\documentclass[aps,prfluids,preprint,groupedaddress]{revtex4-2}
\usepackage{gensymb}
\usepackage{amssymb}
\usepackage{enumitem}
\usepackage{siunitx}
\usepackage[caption=false]{subfig}
\usepackage{graphicx}
\usepackage{xcolor}
\usepackage{soul}

\begin{document}

%\preprint{}

\title{A Mechanistic Pore-Scale Analysis of the Low-Salinity Effect\\
in Heterogeneously Wetted Porous Media
}


\author{Michael G. Watson}
\email[Corresponding author: ]{michael.watson@sydney.edu.au}
%\thanks{}
\affiliation{School of Mathematics and Statistics, University of Sydney, NSW 2006, Australia}

\author{Steven R. McDougall}
%\email[]{}
%\thanks{}
\affiliation{Institute of Petroleum Engineering, Heriot-Watt University, Edinburgh, EH14 4AS, Scotland}


\date{\today}

\begin{abstract}
Over the last two decades, the enhanced oil recovery technique of low-salinity (LS) waterflooding has been a topic of substantial interest in the petroleum industry. Many studies have shown that LS brine injection can increase oil production relative to conventional high-salinity (HS) brine injection, but contradictory results have also been reported and a complete understanding of the underlying mechanisms remains elusive. We have recently developed an innovative, steady-state pore network model to simulate oil recovery by LS brine injection in uniformly wetted pore structures (Watson et al., Transp. Porous Med. \textbf{118}, 201--223, 2017), and we extend this approach here to investigate the mechanisms of the low-salinity effect (LSE) in \textit{heterogeneously} wetted media. We couple a model of capillary force-driven fluid displacement to a novel tracer algorithm and track the evolving salinity front in the pore network as oil and HS brine are displaced by injected LS brine. The wettability of the pore structure is modified in regions where the water salinity falls below a critical threshold, and simulations show that this can have significant consequences for oil recovery.

For networks that contain spanning clusters of both water-wet and oil-wet (OW) pores prior to flooding, our results demonstrate that the OW pores contain the only viable source of incremental oil recovery by LS brine injection. Moreover, we show that a LS-induced increase in microscopic sweep efficiency in the OW pore fraction is a necessary, but not sufficient, condition to guarantee additional oil production. Simulations further suggest that the fraction of OW pores in the network, the average network connectivity and the initial HS brine saturation are key factors that can determine the extent of any improvement in oil recovery in heterogeneously wetted networks following LS brine injection. This study clearly highlights the fact that the mechanisms of the LSE can be markedly different in uniformly wetted and non-uniformly wetted porous media.
\end{abstract}

%\keywords{}

\maketitle

\section{Introduction \label{Intro}}
As the global demand for energy continues to grow, new and innovative enhanced oil recovery (EOR) methods are required to maximise oil production from hydrocarbon reservoirs. One method that has attracted increasing interest over the last decade is low-salinity (LS) waterflooding, which has been shown to be capable of improving oil recovery at the core-scale and the field-scale in both secondary and tertiary modes \citep{Webb04, Gama11, Maha11}. Secondary LS waterflooding refers to the injection of LS brine as an initial method of oil displacement, while tertiary LS waterflooding refers to the injection of LS brine after the use of a conventional secondary displacement technique such as high-salinity (HS) waterflooding. The improvement in oil recovery that is associated with LS brine injection --- the so-called low-salinity effect (LSE) --- is widely believed to arise from a favourable shift in the wettability of the crude oil/brine/rock (COBR) system \citep{Berg10}. However, injection of LS brine does not always increase oil production, and the precise reasons for this patchy and seemingly inconsistent performance remain poorly understood.

The phenomenon of improved oil recovery by injection of LS brine has been known for many decades \citep{Mart59, Bern67}, but oil industry interest has been piqued only recently following the publication of a series of sandstone coreflooding studies by Morrow and colleagues that explored the implications of varying the composition of injected and connate brines \citep{Jadh95, Yild96, Tang97, Morr98, Tang99}. These studies demonstrated strong potential for LS brine injection as an EOR method and suggested that the presence of both clays and connate water, as well as exposure of the rock surface to crude oil, are necessary conditions to achieve a positive LSE. A large number of studies for both sandstones and carbonates have since been published that show an array of positive \citep{McGu05, Ligt09, Ashr10, Yous10, Sheh14, Xie15}, negligible \citep{Zhan06, Rive10, Fjel12} and even negative \citep{Sand11} effects of LS brine injection on oil recovery in secondary and tertiary modes. Research efforts to understand the microscopic mechanisms of the LSE have recently increased \citep{Maha15, Bart17}, but a consistent explanation for the wide scatter in outcomes from coreflooding studies has yet to fully emerge.

Several potential mechanisms for the LSE have been proposed in the literature (see \citet{Morr11}, \citet{Skau13} and \citet{Shen14} for comprehensive reviews on this topic). Destabilisation and migration of fines is a frequently reported consequence of LS brine injection that may influence oil recovery by mobilising oil that is attached to these oil-wet (OW) clay particles \citep{Tang99}. Subsequent deposition of mobilised fines in small downstream pores may also lead to microscopic diversion of fluids by increasing pressure gradients elsewhere in the network \citep{Spil12}. Other LS mechanisms that have been proposed involve disturbances to the established chemical equilibrium of the COBR system, such as pH variation \citep{Aust10}, multicomponent ion exchange \citep{Lage06} and electrical double layer expansion \citep{Ligt09, Maha15}. Given the number of different mechanisms that have been proposed (and the often contradictory results obtained from coreflooding studies), it seems plausible that the LSE actually represents the manifestation of several factors that act in concert. Whilst this may be true, there remains a general acknowledgement that the overall \textit{consequence} of LS brine injection is a shift in the rock wettability, and that this shift is most likely towards increased water-wetness or decreased oil-wetness \citep{Agba09, Nasr11}. Direct evidence for wettability alteration has recently been provided by \citet{Khis17}, who used micro-CT imaging to visualise a reduction in oil-water contact angles from an average of 115{\degree} to an average of 89{\degree} inside miniature core samples following LS brine injection. This change in wettability led to an overall increase in oil recovery, which was explained by increased water invasion of small- and medium-sized pores due to their reduced capillary entry pressure thresholds. Note that here, and in all that follows, the oil-water contact angles in individual pores have been measured through the aqueous phase. By convention, we use the term water-wet (WW) to refer to pores that have equilibrium contact angles less than 90{\degree} and OW to refer to pores that have equilibrium contact angles greater than 90{\degree}. The strength of the wetting preference in a given pore is determined by the proximity of its contact angle to 90{\degree} (e.g.\ a strongly water-wet pore has contact angle close to 0{\degree} and a weakly water-wet pore has contact angle close to 90{\degree}).

Pore network models have been widely applied in the oil industry to study pore-scale fluid displacement phenomena in a variety of contexts (see \citet{Blun01}, \citet{Joek12} and \citet{Blun17} for excellent reviews of this work). Following on from the theoretical approach developed by \citet{Sorb10}, we have recently developed \textit{steady-state} and \textit{unsteady-state} pore network modelling approaches to investigate the pore-scale displacement behaviour that arises from LS-induced wettability modification in \textit{uniformly wetted} networks \citep{Wats17, Bouj18}. In these studies, we have used tracer algorithms to track the evolution of water salinity when LS brine is injected to displace oil from \textit{in silico} networks that also contain HS brine. Oil-water contact angles of individual pores were modified when local brine salinity fell below a critical threshold, and the results have indicated that this can have a dramatic impact on both the sequence of pore-filling and the overall volume of oil recovered from the network.

The fluid displacement methodologies in steady-state and unsteady-state pore network models are distinctly different, and this has important implications for the spatio-temporal tracking of water salinity in LS brine injection simulations. In steady-state models (such as \citet{Wats17} and the current study), viscous forces are assumed to be negligible and the displacement of oil from the network is driven purely by capillary forces between the oil and the injected brine. This approach is most applicable to far-field regions of the oil reservoir (i.e.\ the bulk of the reservoir away from injection and producing wells). The capillary pressure $P_c$ (which we define as the pressure difference between the oil and water phases, $P_c = P_o - P_w$) is reduced in a stepwise manner and oil is displaced from several pores at each stage of the simulation. Saturation changes in the network in steady-state models are therefore distinct and governed solely by the capillary entry thresholds of the pores. Hence, in steady-state LS waterflooding models, the evolution of brine salinity cannot be tracked explicitly and must be estimated \textit{a posteriori}. In unsteady-state models (such as \citet{Bouj18}), the displacement of oil from the network is driven by both viscous and capillary forces (note, however, that the underlying methodology is equally applicable at low water injection rates where viscous forces may be negligible). For unsteady-state models, a fixed rate of water injection is typically assigned at the network inlet. Individual oil-water menisci are then explicitly tracked as they move through each pore, and the simulation proceeds by completely filling one single pore with water at each timestep. Unsteady-state models are therefore much more computationally expensive than steady-state models and this is prohibitively restrictive for 3D simulation.

Much like the experimental coreflooding studies discussed above, our recently published steady-state and unsteady-state LS waterflooding models have demonstrated a range of outcomes that include positive, neutral and negative effects of LS brine injection. \citet{Bouj18} performed unsteady-state LS brine injection simulations on uniformly wetted 2D pore networks and showed that the outcome of dynamic contact angle reduction by LS brine can be influenced by parameters such as the water injection rate and the oil-water viscosity ratio. Results showed that dynamically reducing capillary forces in the network can lead to a marked change in the displacement regime (e.g.\ from capillary fingering to stable displacement, or from capillary fingering to viscous fingering) and that this can ultimately result in either a positive or negative LSE. \citet{Bouj18} also found that the particular combination of initial and LS-modified wettability (e.g.\ strongly OW to weakly OW, strongly OW to neutral wet) can be an important factor. This finding is consistent with the results of \citet{Wats17}, who performed steady-state simulations of LS brine injection in uniformly wetted 3D pore networks. \citet{Wats17} explored a range of possibilities for the combination of initial and LS-modified wettability, including scenarios where LS brine caused contact angles to \textit{increase} --- indeed, the phenomenon of increased oil-wetness following LS brine injection has been reported on several occasions \citep{Buck97, Sand11, Fjel12, Shak13} and our results indicated that this wettability shift \textit{can} lead to increased oil production, particularly if LS brine causes contact angles to change from WW to OW. \citet{Wats17} also demonstrated that the pore size distribution (PSD) of the network can be a key factor in determining the extent of any LSE. In particular, for networks with pore radii uniformly distributed between $R_{min}$ and $R_{max}$, we showed that the potential for LS brine injection to improve oil recovery by reducing pore contact angles increases as the ratio $R_{max}/R_{min}$ decreases.

In more general terms, the key insight from the study of \citet{Wats17} was the identification of two distinct pore-scale effects that can be used to explain the outcome of dynamic LS-induced contact angle modification in a particular scenario. First, the \textit{``pore sequence effect''}, which is characterised by an \textit{overall change in the size distribution of pores displaced} by injected LS brine, and second, the \textit{``microscopic sweep efficiency effect''}, which is characterised by an \textit{overall change in the total fraction of pores displaced} by LS brine. Importantly, we showed that these two effects can act independently or synergistically depending on the choice of parameters, and also that both can have positive and negative consequences for overall oil recovery following LS brine injection.

The studies by \citet{Wats17} and \citet{Bouj18} have provided many new insights into the pore-scale displacement phenomena that can accompany dynamic wettability modification and demonstrate the potential of pore network modelling as a powerful and inexpensive tool to help interpret the results of experimental LS coreflooding studies. However, the results from these studies were obtained by considering only \textit{in silico} networks that initially contain either 100\% WW or 100\% OW pores. Such idealised distributions of wettability are unlikely to hold for many of the rock samples utilised in experimental studies, where a combination of heterogeneous mineralogy and core preparation techniques (e.g.\ aging) will lead to a mixture of wettability conditions within the core. Therefore, in this paper, we extend the steady-state modelling approach of \citet{Wats17} to investigate the LSE in networks with \textit{non-uniform} distributions of initial wettability. Our results suggest that the conditions required to achieve a positive LSE in non-uniformly wetted networks may be much more stringent than for uniformly wetted networks, and, as will be seen, depend on parameters that we have not reported to date. We will begin in the next section by restating our modelling methodology in full, and we will then present a series of simulation results and parameter sensitivities in Section~\ref{Results}. We will provide an in-depth discussion of the implications of our results in Section~\ref{Disc}, before concluding with a broad summary of the outcomes of the study in Section~\ref{Conclu}.

\section{Model Methodology \label{Model}}

\subsection{Pore Network Initialisation \label{PNInit}}
We stress from the outset that our goal in this paper is not to make quantitative predictions, but rather to use pore network modelling techniques to gain new, qualitative insights into the mechanisms of the LSE in networks of non-uniform wettability. Therefore, as in our previous studies of LS brine injection in uniformly wetted networks \citep{Wats17, Bouj18}, we choose to represent the pore space as a simple 3D network of interconnected capillary elements --- more complex network architectures with irregular distributions of connectivity and pore geometry could be considered, but we maintain an idealised approach at this stage to better highlight the underlying pore-scale behaviours and to simplify the analysis of our results. Hence, we assume that, in general, the properties of the individual pores in our networks can be described using the scaling laws proposed by \citet{McDo95}. In this theory, a capillary entry radius $R$ can be assigned to each pore, and the pore volume $V$ and pore conductance $g$ can then be described by the proportionalities $V\left(R\right) \propto R^{\nu} \left(0 \leqslant \nu \leqslant 2\right)$ and $g\left(R\right) \propto R^{\mu} \left(2 \leqslant \mu \leqslant 4\right)$, respectively. We shall initially assume that our pore elements are perfectly cylindrical (i.e.\ $\nu = 2$ and $\mu = 4$), but we will later reduce the value of $\nu$ to give a more representative distribution of individual pore volumes across the network. Note that the assumption of cylindrical pore elements precludes the possibility of explicit film and corner flow modelling in our networks. As discussed extensively in both \citet{Wats17} and \citet{Bouj18}, wettability alteration by LS brine transported in the corners of angular pores could have additional implications for oil recovery. We will focus here simply on bulk transport mechanisms but note that we are currently exploring the consequences of LS brine transport in films and will report this work in a future publication.

\begin{figure}
\subfloat[\label{WetClassFiga}]{{\includegraphics[width=5.2cm]{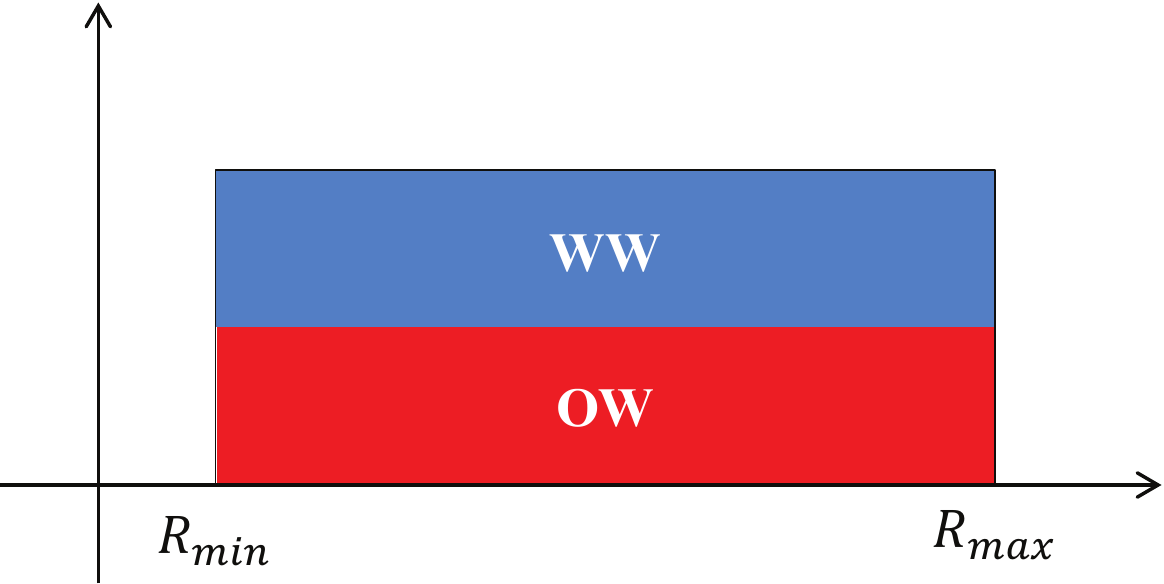}}}
\quad
\subfloat[\label{WetClassFigb}]{{\includegraphics[width=5.2cm]{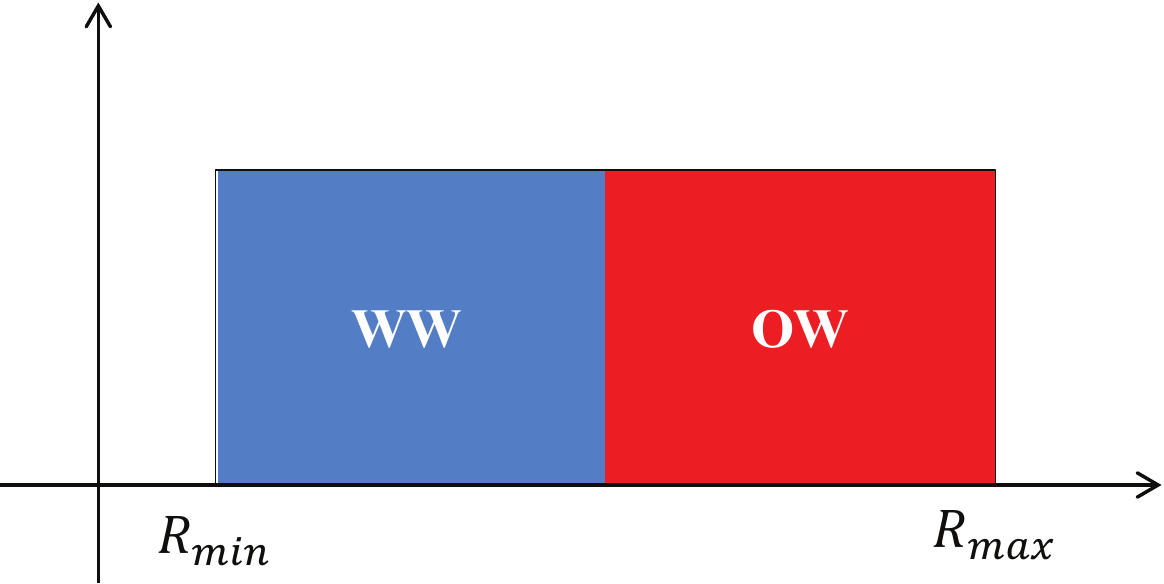}}}
\quad
\subfloat[\label{WetClassFigc}]{{\includegraphics[width=5.2cm]{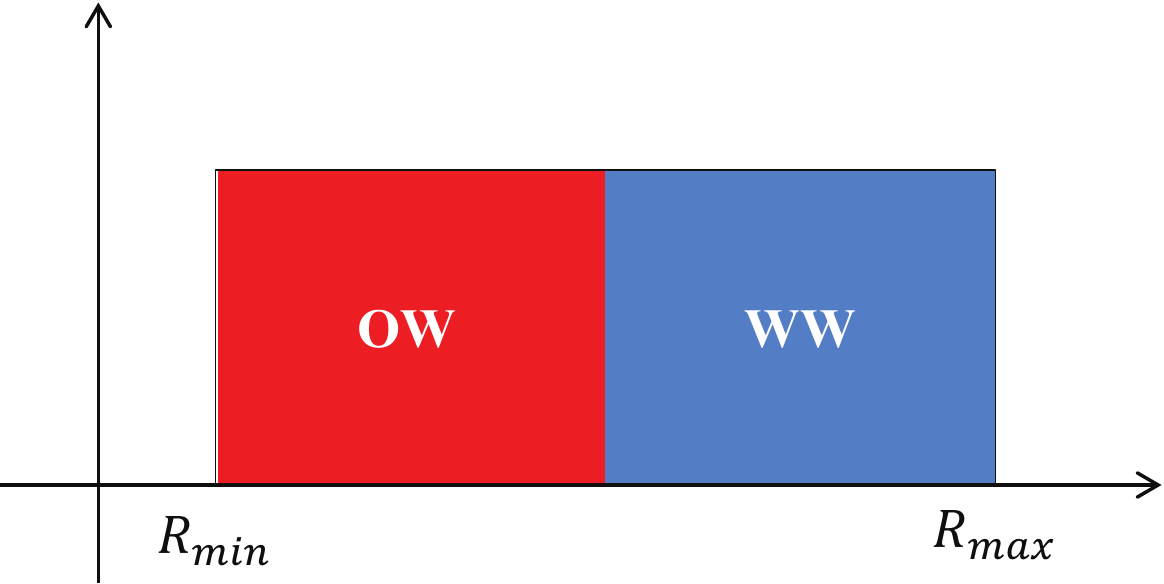}}}
\caption{Schematic plots of the WW and OW pore size distributions in (a) FW, (b) MWL and (c) MWS systems. \label{WetClassFig}}
\end{figure}

For each simulation, we generate a 3D pore space by defining the network dimensions (i.e.\ the number of nodes in each direction), the average pore length $L$, and the network co-ordination number $\bar{Z}$, which determines the average number of capillary elements (pores) connected to each junction (node). Individual pores are assigned both a radius $R$ from a chosen pore size distribution (PSD) and a contact angle $\theta$ that defines the initial wettability of the pore. Of course, in a real porous medium, we might expect a certain level of intra-pore wettability variation due to localised heterogeneity in rock mineralogy and pore geometry. However, wettability variation at the scale of a single pore would be difficult to implement in any meaningful way and, moreover, this type of data is rarely measured experimentally. We therefore assume that $\theta$ represents an \textit{effective} contact angle that captures some spatial average of the wetting properties in each pore. In this paper, we are only concerned with networks that initially contain both WW and OW pores, so we introduce the parameter $\alpha$ to represent the fraction of pores that are initially assigned OW contact angles. If WW and OW contact angles are assigned to pores at random (i.e.\ with no correlation between pore size and wettability), the network is termed fractionally-wet (FW). However, in addition to FW networks, there also exist two well-recognised wettability classes that display a distinct correlation between pore size and pore wettability. These are the mixed-wet large (MWL) class, where only the \textit{largest} pores are OW, and the mixed-wet small (MWS) class, where only the \textit{smallest} pores are OW (Figure~\ref{WetClassFig}). See \citet{Skau07} for an interesting discussion regarding these three wettability classes in the context of related experimental work. Here, we will mostly use networks of the MWL class, but, where possible, we will also interpret the implications for similar FW and MWS networks.

\subsection{Steady-State Oil Displacement \label{SSDisp}}
The steady-state approach to waterflood simulation is well-established and has been reported previously \citep{Blun97, Dixi99}. Briefly, water is introduced at the inlet face of an oil-bearing network and a stepwise reduction of the $P_c$ allows water invasion of oil-filled pores with sequentially decreasing capillary entry pressures. As we are neglecting film flow at this stage, water ingress can only occur via piston-like displacement of pores with a bulk water connection to the network inlet --- this takes place at capillary entry pressure $2\sigma\cos\left(\theta\right)/R$, where $\sigma$ denotes the oil-water interfacial tension. In a heterogeneously-wet system, a representative HS flood proceeds as follows: (i) the flood begins at a high, positive $P_c$ and water first invades the smallest WW pores that are connected to the inlet; (ii) the displacement continues at positive (but decreasing) $P_c$, filling successively larger WW pores until all displaceable WW pores have been invaded; (iii) the water then fills the largest OW pores that are connected either to the inlet or to bulk water, and proceeds to displace oil from progressively smaller OW pores as the $P_c$ becomes increasingly negative. Oil that becomes isolated from the network outlet by invading water is trapped and remains immobile. Note that we do not consider snap-off of oil-filled pores due to notional water film accumulation in WW pores, nor do we allow oil to drain from the network through OW pathways. Simulations proceed according to the above rules until no further oil can be displaced from the network.

Note that, in this stepwise oil displacement procedure, the simulated rate of $P_c$ decline can be regarded as an implicit measure of the physical waterflood injection rate. Under low water injection rate conditions (small $P_c$ steps), displacement of oil by water will occur pore-by-pore. Conversely, at higher injection rates (large $P_c$ steps), many capillary entry thresholds will be satisfied at the same time, and water will displace oil from many pores simultaneously. Hence, different injection rates ($P_c$ step sizes) result in different pore displacement patterns.

\subsection{Tracer Injection Algorithm \label{Tracer}}
The steady-state LS waterflooding model uses an innovative approach to track the evolution of water salinity during oil displacement from the pore network. In this section, we provide a complete description of our salinity tracking methodology, followed by a detailed physical interpretation of our modelling assumptions.

\subsubsection{Methodology \label{TracerMethod}}
Following \citet{Wats17}, we track the dispersion of HS and LS brines in the network by coupling the steady-state oil displacement model to an unsteady-state tracer injection algorithm. We assume that HS brine (tracer concentration of zero) and LS brine (tracer concentration of one) are miscible and that they instantaneously mix in pore elements. The normalised brine salinity in a general water-filled pore is therefore given by 1 -- (\textit{tracer concentration}). During periods of HS brine injection, there is no need to track the water salinity in the network and simulations proceed as detailed in Section~\ref{SSDisp}. However, when LS brine is injected into a network that contains HS brine, the tracer algorithm is applied after each saturation change to track the dispersion of the newly-introduced fluid. In the absence of explicit, dynamic information about the LS brine influx, we effectively ``rewind'' time to before the most recent saturation change and use the tracer approach to estimate new brine salinities in the water-filled pores. Contact angles of oil-filled pores are then updated according to their local brine salinity (see Section~\ref{TC2CA}), and this localised wettability modification can alter the sequence of pore-filling at subsequent displacement steps.

\begin{figure}
\includegraphics[width=12.9cm]{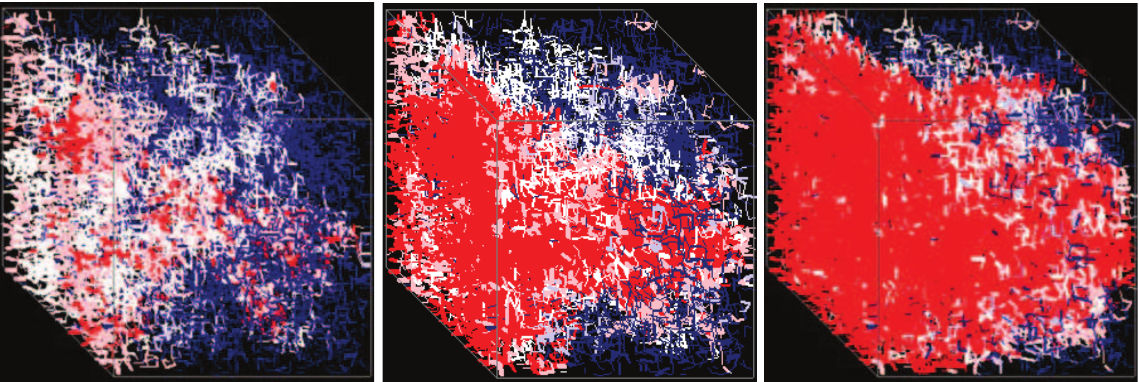}
\caption{Snapshots of an evolving salinity front during a typical model simulation where injected LS brine (salinity = 0) displaces oil from a pore network that contains HS brine (salinity = 1). Colours of individual pores correspond to normalised salinity values as follows: \textit{red} [0, 0.25); \textit{pink} [0.25, 0.5); \textit{white} [0.5, 0.75); \textit{light blue} [0.75, 1); \textit{dark blue} 1. \label{SalFrontFig}}
 \end{figure}

In this paper, we consider two different modes of LS waterflooding: (1) a secondary mode where LS brine is injected into a network that contains an initial (spatially-distributed) HS brine saturation $S_{wi}$; and (2) an early tertiary mode where LS brine injection commences at water breakthrough following a period of HS brine injection ($S_{wi} = 0$). Our simulations therefore involve both pre-breakthrough and post-breakthrough LS brine injection, and we use different methods to update the salinities of water-filled pores in each of these two scenarios. Prior to water breakthrough, the volumes of HS and LS brine that occupy each water cluster are known exactly and we estimate the water salinity in each pore using a cluster-scale averaging algorithm. After breakthrough, however, the resident HS and LS brines disperse due to flow and estimation of new water salinities requires a more detailed calculation. See Figure~\ref{SalFrontFig} for some images of the salinity evolution during a typical simulation as oil and HS brine are displaced from a pore network by LS brine injection.

In the case of pre-breakthrough LS brine injection, the tracer concentrations in water-filled pores are updated after each saturation change according to the following steps:
\begin{enumerate}[label=(\alph*)]
\item Identify each unique inlet-connected water cluster.
\item For each cluster, determine the total cluster volume $V_{clus}$, the total mass of tracer newly added to the cluster $M_{new}$ (i.e.\ pores just displaced by injected LS brine) and the previous mass of tracer in the cluster $M_{clus}$ (i.e.\ pores displaced by HS or LS brine at earlier times).
\item \begin{sloppypar} Assign to each pore in the cluster the averaged tracer concentration $C_{clus} = \left(M_{new}+M_{old}\right)/V_{clus}$. \end{sloppypar}
\end{enumerate}

For post-breakthrough LS brine injection, we use a two-stage process to calculate the new tracer concentrations associated with the latest saturation change in the network. First, we assign initial tracer concentrations to all newly displaced pores by estimating the salinity of the water driven into them by the invading LS brine. We then simulate a period of tracer flow in the spanning water cluster, where the total injection time accounts for the fact that the volume of brine injected to achieve the latest saturation change will typically be larger than the volume of oil produced (i.e.\ some of the injected LS brine will simply bypass the oil and flow out of the network). Note that we assume a Poiseuille flow law in each pore element for all tracer flow calculations.

The salinity assignment procedure at each Pc step for post-breakthrough LS brine injection can be summarised as follows:
\begin{enumerate}[label=(\arabic*)]
\item For an arbitrary pressure drop between the inlet and outlet faces of the network, calculate the global pressure solution in the spanning water cluster and determine the individual pore flow rates. Rescale this solution to achieve the desired fixed injection rate $Q$, and determine the associated pressure drop $\Delta P_{new}$.
\item Calculate the volume of oil $\Delta V$ just displaced from the network.
\item Estimate the time $T_{\Delta V}$ required to achieve this oil displacement, according to the equation:
\begin{equation} 
	T_{\Delta V} = \frac{\Delta V}{Q}\left[\frac{\Delta P_{old}}{\Delta P_{old} - \Delta P_{new}}\right], \label{TDVEq}
\end{equation}
where $\Delta P_{old}$ denotes the pressure drop from the previous displacement step. If $T_{\Delta V}$ exceeds the time $T_{PV}$ required to inject one whole pore volume of fluid, set $T_{\Delta V} = T_{PV}$.
\item Traverse flow pathways through the spanning water clusters and assign to all newly displaced and newly connected pores the flow-weighted average tracer concentration of their upstream neighbours.
\item Follow steps (a)--(c) described earlier for all non-spanning inlet-connected water clusters and for all ``dead-end'' water clusters (i.e.\ pores that branch from a flowing water cluster but do not yet support flow themselves). For dead-end clusters, the source concentration for new mass $M_{new}$ is the flow-weighted average tracer concentration in the upstream neighbours of the node from which the cluster branches.
\item Calculate the total mass of tracer $M_0$ added to pores in steps (4) and (5), and calculate the time $T_0$ required for this period of pre-assignment according to $T_0 = M_0/Q$. Simulate dynamic tracer injection in the flowing water-filled pores for the remaining period of time $T^* = T_{\Delta V} - T_0$ (see the Appendix for a fuller description of the methodology).
\end{enumerate}

Equation~(\ref{TDVEq}) is a crucial component of the tracer assignment procedure, and its format has been chosen to reflect qualitative observations from experimental coreflooding studies. The term $\Delta V/Q$ quantifies the \textit{minimum} length of time $T_{min}$ required for the most recent network saturation change, while the viscous pressure ratio term provides a modulating factor that determines the extent to which this minimum time is lengthened by water egress from the network. If the reduction in the pressure drop between successive $P_c$ steps is large (e.g.\ when a percolating water cluster becomes more connected), we assume that minimal water will be lost to the outlet during the oil displacement ($T_{\Delta V}$ not much larger than $T_{min}$). However, if only a small reduction in the pressure drop can be achieved (e.g.\ when the water phase has a well-established spanning cluster), we assume that the process of oil displacement will be highly inefficient ($T_{\Delta V}$ significantly larger than $T_{min}$). Note that the choice of water injection rate $Q$ provides a physical time scaling for the tracer injection.

We briefly comment on the limiting criterion in step (3) above and emphasise that, in practice, the condition $T_{\Delta V} > T_{PV}$ is very rarely satisfied. Moreover, when this condition is satisfied, it is always towards the end of a simulation when little untrapped oil remains, and the global pressure drop in the water phase becomes insensitive to further small saturation changes in the network. The inclusion of the limiting criterion therefore has no real consequences for the simulation results, and we include it in the model simply to prevent excessively long run times in the circumstances just described.

\subsubsection{Physical Interpretation \label{TracerInterp}}
Based on the tracer injection methodology described above, it should be clear that the outcome of a given LS waterflooding simulation will intrinsically depend on the number of $P_c$ steps that are taken to displace the resident oil. This is a deliberate feature of the model and reflects the fact that the $P_c$ step size is correlated to the physical waterflood injection rate. In any given simulation, oil is displaced from the network by discretely reducing the $P_c$ from some initial (maximum) value to some final (minimum) value. If the $P_c$ is reduced over many \textit{small steps} (mimicking a low water injection rate), the tracer injection algorithm would predict an \textit{inefficient} displacement of the oil after water breakthrough (i.e.\ an injected water volume \textit{much larger} than the produced oil volume over the course of a simulation). On the other hand, if the $P_c$ is reduced over few \textit{large steps} (mimicking a higher water injection rate), the algorithm would predict an \textit{efficient} displacement of the oil after water breakthrough (i.e.\ an injected water volume only \textit{slightly larger} than the produced oil volume over the course of a simulation).

For all simulations performed in this study, the $P_c$ has been stepped from its maximum value to its minimum value in no more than 100 steps (mirroring an intermediate water injection rate and providing a sensible base case scenario). As a simulation proceeds, each new $P_c$ is chosen to ensure that the entry pressures of approximately 1\% of the oil-filled pores in the network are newly satisfied. If more than 1\% of the oil-filled pores have become newly accessible due to LS-induced contact angle modification at the previous simulation step, the $P_c$ is maintained at its existing level for the subsequent simulation step. For 100 steps, we find that our simulations require around 2--4 PVs of water injection to remove all of the displaceable oil from the network. A total injection volume of 2--4 PVs is broadly consistent with the volume of water required to displace the resident oil from core samples in experimental studies.

The tracer injection algorithm detailed in Section~\ref{TracerMethod} combines both advection modelling and a cluster-scale averaging procedure to track the evolution of brine salinity in the aqueous phase. A physical and methodological justification for this hybrid approach is given below, where we interpret the implications of our assumptions in the context of both tertiary and secondary LS brine injection. In tertiary LS simulations, we initially consider cases where $S_{wi} = 0$ and mixing of HS and LS brines occurs only after injected HS brine has broken through. At breakthrough, the HS brine in the network will be mostly distributed in one or two spanning clusters and several large (inlet-connected) non-spanning clusters. In the spanning water cluster(s), and any others that subsequently become spanning, tracer evolution is determined by means of the global flow solution. Here, we consider only advective mixing of the HS and LS brines, but this is reasonable because the frontal advance rate in a typical coreflood (1--5 m/day) should limit the capacity for diffusive mixing in the tortuous pore space. In the non-spanning water clusters that grow (or emerge from the inlet) by LS brine injection, we assume perfect (instantaneous) mixing of the HS and LS brines. While the assumption of perfect mixing (rather than advective mixing) of HS and LS brines in non-spanning water clusters seems restrictive, note that, in any non-spanning water cluster that contains mostly HS brine or mostly LS brine, perfect mixing should make little difference to the salinity in most pores in the cluster. If a non-spanning water cluster instead contains similar volumes of HS and LS brine, the perfect mixing assumption is less valid and (relative to advective mixing) one would anticipate a slight reduction in contact angle modification events proximal to the network inlet. Accurate tracking of LS ingress in non-spanning water clusters requires a more involved unsteady-state modelling approach, such as that reported by \citet{Bouj18} for uniformly wetted systems.

In secondary LS simulations, we include cases where LS brine invades a network that is initially populated by randomly distributed clusters of HS brine. We again assume perfect mixing of HS and LS brines in non-spanning water clusters. Therefore, as clusters of LS brine emerge from the network inlet, they immediately assimilate with the connate HS brine clusters that they contact. In reality, this assimilation process would be a local phenomenon, whereby each HS brine cluster that is contacted by LS brine is diluted locally by convective mixing with its neighbouring LS brine-filled pores. However, as LS brine clusters develop within the tortuous pore space, connate HS brine clusters are contacted around the entire perimeter of each LS brine cluster (this can be envisaged as a series of fingered LS brine structures that are ``studded'' with small HS brine clusters). Consequently, the assumption that HS brine clusters undergo perfect mixing with their host LS brine cluster(s) at each $P_c$ step is not as restrictive as it may first appear.

We further remark that while we have neglected to explicitly consider mechanisms of salinity dispersion, we surmise that dispersive effects would play a secondary role compared to the first-order impacts of network topology and wettability distribution during advective water transport. Phenomena such as Taylor dispersion and diffusion within dead-end loops could be included in a more refined future model, and these mechanisms may well provide additional insights into the complex LS waterflooding process.

\subsection{Coupling Tracer Concentration to Contact Angle \label{TC2CA}}
After appropriate tracer concentrations have been assigned to water-filled pores, we use the updated salinity map in the network to locally modify the contact angles of remaining oil-filled pores. In the absence of definitive data on the relationship between brine salinity and the pore-level wettability change within rock structures, we adopt a modest wettability modification methodology that reflects key qualitative observations from LS coreflooding experiments. Results from laboratory studies suggest that the LSE will be negligible for injected brine salinities above approximately 5000 ppm \citep{Lage06}, but may become increasingly efficacious at lower salinity \citep{Tang97, Ashr10}. An expression that captures increasing contact angle modification with increasing tracer concentration therefore seems reasonable, and to simplify the analysis of our results we propose the following Heaviside model to describe the dynamic modification of contact angles in the network:
\begin{equation}
  \theta=\left\{
  \begin{array}{@{}ll@{}}
    \theta_{HS}, & C_N < C^* \\
    \theta_{HS} - \Delta\theta, & C_N \geqslant C^*.
  \end{array}\right. \label{ThetaEq}
\end{equation} 
Here, $\theta_{HS}$ denotes the unmodified contact angle of an oil-filled pore, $\Delta\theta$ quantifies the extent of LS-induced contact angle modification, $C_N$ denotes the maximum tracer concentration in all neighbouring water-filled pores and $C^*$ denotes the critical tracer concentration required for contact angle modification. Equation~(\ref{ThetaEq}) effectively assumes that when brine of sufficiently low salinity invades a pore, the brine will contact the oil/rock interface of neighbouring oil-filled pores and locally alter the surface wettability at the neighbouring pore entrances. If the capillary entry pressure of any neighbouring pore is then satisfied, LS brine will enter the pore and there is an implicit assumption that the wettability of the entire host pore surface will be progressively altered as the brine invades. We assume that the modification of wettability in each pore arises from an appropriate chemical reaction (e.g.\ multi-component ion exchange or electrical double layer formation). However, as this reaction operates on a much faster timescale than the time between successive $P_c$ steps, we do not explicitly model the associated chemical mechanisms.

\section{Results \label{Results}}
\begin{figure}
\includegraphics[width=8cm]{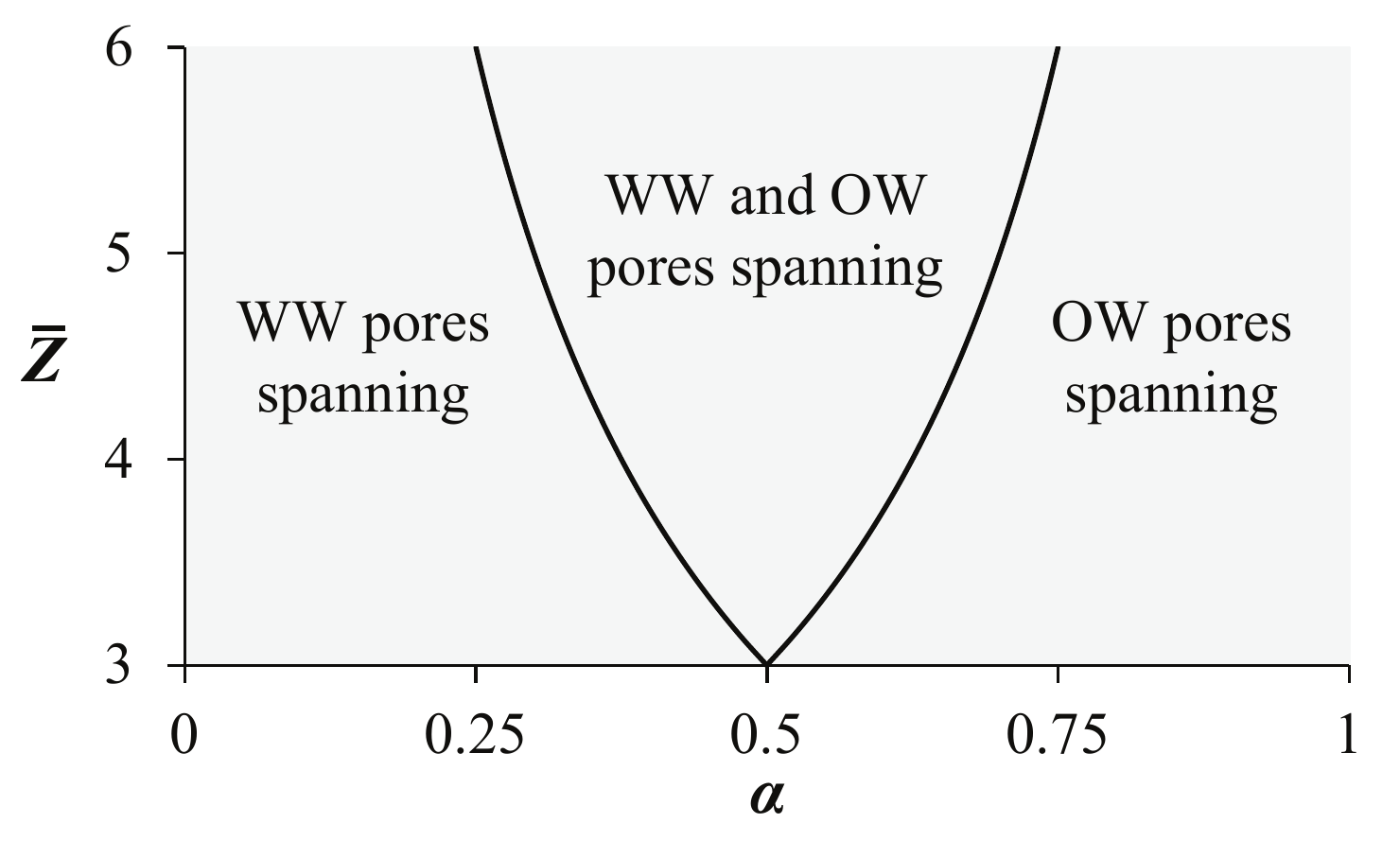}
\caption{Plot that shows how the number of spanning wettability clusters in a regular 3D pore network is determined by the fraction of OW pores in the network ($\alpha$) and the average network co-ordination number ($\bar{Z}$). In the left-hand region of the plot only WW pores span the network, in the right-hand region of the plot only OW pores span the network, and in the central region both WW and OW pores span the network. The curves that delineate these regions are derived using results from percolation theory (see equation~(\ref{PercEq}) in the main text). \label{PercFig}}
\end{figure}

As discussed in Section~\ref{Intro}, we have previously used the above modelling approach to investigate pore-scale mechanisms that potentially underlie the LSE (or lack of) in networks of \textit{uniform} initial wettability (i.e.\ networks that consist of either 100\% WW pores or 100\% OW pores prior to LS flooding; \citet{Wats17}). The results of this earlier study, which explored the implications of dynamic wettability alteration by LS brine injection, revealed a suite of potential modifications to pore filling sequences that can offer some insight into the apparent inconsistencies that have been reported in the experimental LS waterflooding literature. To obtain a more complete understanding of the pore-scale displacement phenomena associated with LS brine injection, the current study investigates the consequences of dynamic, LS-induced wettability alteration in networks of \textit{non-uniform} initial wettability. Note that by non-uniform we specifically refer to networks that initially possess \textit{spanning clusters of both WW and OW pores}. Percolation theory therefore stipulates that we require:
\begin{equation} 
	\frac{D}{\bar{Z}\left(D - 1\right)} \leqslant \alpha \leqslant 1 - \frac{D}{\bar{Z}\left(D - 1\right)}, \label{PercEq}
\end{equation}
where $D$ denotes network dimensionality. Figure~\ref{PercFig} shows a plot of the parameter combinations $\left(\alpha, \bar{Z}\right)$ that satisfy this condition for regular 3D pore networks: note that, for networks that contain both WW and OW pores but only one spanning wettability cluster, our model produces results that can be interpreted from the uniform wettability study performed in \citet{Wats17}.

Although there is a commonly held view in the experimental literature that LS waterflooding can lead to increased water-wetness in COBR systems, a LS-induced increase in oil-wetness has also been reported on several occasions \citep{Fjel12, Shak13}. In our previous study with uniformly-wetted networks, these contrasting reports motivated us to explore a range of possible scenarios by assuming that LS brine could lead to wettability modification towards either increased water-wetness or increased oil-wetness. However, given the additional complexity introduced by using networks with non-uniform initial wettability, in the current study we shall focus predominantly on cases where LS injection causes initially WW pores to become more WW and initially OW pores to become less OW. The implications for other possible scenarios, including those where LS brine causes pores to change their wetting class (i.e.\ OW to WW or vice versa), will be summarised in Section~\ref{Disc}. For the various scenarios, we will also consider how the LSE is likely to vary according to the wettability classification of the underlying pore network (i.e.\ FW, MWL or MWS).

In \citet{Wats17}, we adopted a LS injection methodology that lay somewhere between secondary and tertiary in that we captured elements of both approaches by assuming the injection of HS brine until water breakthrough and the injection of LS brine thereafter. While we return to this protocol for the initial part of the current study, our findings suggest that a more conventional (and experimentally-relevant) secondary LS injection methodology is required to provide additional important insights into the mechanisms of the LSE in networks of non-uniform wettability. Hence, the final results section is dedicated to secondary simulations, with LS brine introduced into a HS-bearing network at the very outset of the waterflood. Regardless of the particular scheduling of brine injection, for each parameter combination investigated in this study, two separate simulations are performed: one producing a standard steady-state displacement with HS brine injection only, and the other involving LS brine injection with associated wettability modification.  By generating both sets of results, the impact of any alteration to pore filling sequence by LS flooding can be readily assessed.

\subsection{Post-Breakthrough Low Salinity Waterflooding \label{PBLSW}}
For consistency with the approach of \citet{Wats17}, we begin our investigation of heterogeneously-wetted networks by assuming that HS brine is injected until water breakthrough and LS brine is injected thereafter. Recall that we use a tracer algorithm to simulate mixing of this LS water with pre-existing HS water in the network. This leads to a spatio-temporal evolution of salinity within the aqueous phase and contact angle modification in neighbouring oil-bearing pores that satisfy the condition given in equation~(\ref{ThetaEq}). Where applicable, we have used base case parameters that are identical to those used in \citet{Wats17}. Thus, unless otherwise stated, the following properties have been utilised in this section:
\begin{itemize}[label=--]
\item 30 nodes $\times$ 25 nodes $\times$ 25 nodes network (with an average pore length of \SI{333}{\micro\metre});
\item uniform PSD with minimum radius $R_{min} =$  \SI{1}{\micro\metre} and maximum radius $R_{max} =$  \SI{50}{\micro\metre} (note that we deliberately choose a uniform distribution for ease of analysis);
\item pores are perfectly cylindrical: volume scaling $V\left(R\right) \propto R^2$ and conductance scaling $g\left(R\right) \propto R^4$;
\item oil-wet pore fraction $\alpha = 0.5$;
\item average connectivity of pore elements $\bar{Z} = 5$;
\item initial water saturation $S_{wi} = 0$;
\item critical tracer concentration for localised wettability change $C^* = 0.8$ (i.e.\ contact angle modification requires at least a fivefold dilution of resident HS brine);
\item contact angle change due to LSE $\Delta \theta = 20\degree$ (note that this value is consistent with the average 16{\degree} shift reported by \citet{Khis17}).
\end{itemize}

\begin{figure}
\includegraphics[width=8cm]{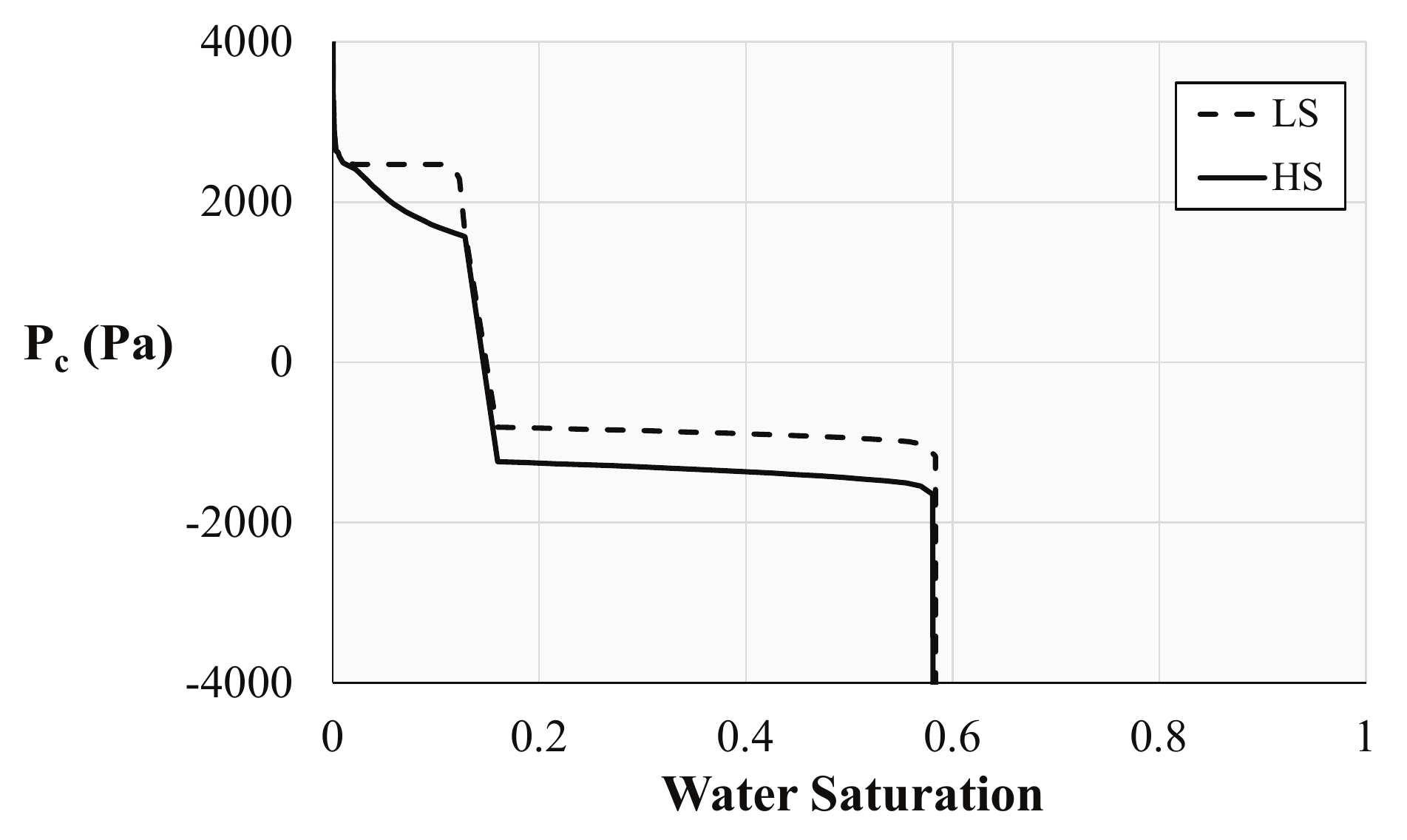}
\caption{Simulated $P_c$ curves for HS brine injection (solid line) and post-breakthrough LS brine injection (dashed line) into a MWL network with 50\% WW pores ($\theta_{HS,WW} = 60\degree$) and 50\% OW pores ($\theta_{HS,OW} = 140\degree$). \label{PBPcFig}}
\end{figure}

We begin by performing waterflood simulations on a MWL network that comprises 50\% moderately WW pores and 50\% strongly OW pores --- the initial contact angles for WW and OW pores are assumed to be $\theta_{HS,WW} = 60\degree$ and $\theta_{HS,OW} = 140\degree$, respectively. Dynamic wettability modification by LS injection, which is achieved when the local tracer concentration exceeds the critical value $C^*$, therefore leads to pores of either stronger water-wetness ($\theta_{LS,WW} = \theta_{HS,WW} - \Delta \theta = 40\degree$) or more moderate oil-wetness ($\theta_{LS,OW} = \theta_{HS,OW} - \Delta \theta = 120\degree$). Note that, while it would be entirely feasible to consider a range of different values for $\theta_{HS,WW}$, $\theta_{HS,OW}$ and $\Delta \theta$ across the pore network, this would only serve to obfuscate the behaviours that we are seeking to understand. The $P_c$ curves obtained from the HS and LS simulations are presented in Figure~\ref{PBPcFig} (recall that HS brine is injected until breakthrough in the latter case, so the two displacements are identical up to that point). From these curves, a number of interesting observations can immediately be made, including: (i) during the respective LS imbibition and drainage cycles, the majority of the oil is recovered at a higher $P_c$ than previously possible; (ii) the shape of the LS $P_c$ curve during the imbibition leg of the waterflood (oil displacement from WW pores) is quite different to that seen with HS injection and features a period of recovery at fixed $P_c$; (iii) the HS and LS $P_c$ curves during drainage (oil displacement from OW pores) appear to differ only by a uniform shift in the $P_c$; and (iv) LS brine injection fails to improve the overall oil recovery at the end-point $P_c$. To fully understand the mechanisms that underlie these results, a thorough investigation of the respective HS and LS pore filling sequences is required.

\begin{figure}
\subfloat[\label{PBFOFiga}]{{\includegraphics[width=8cm]{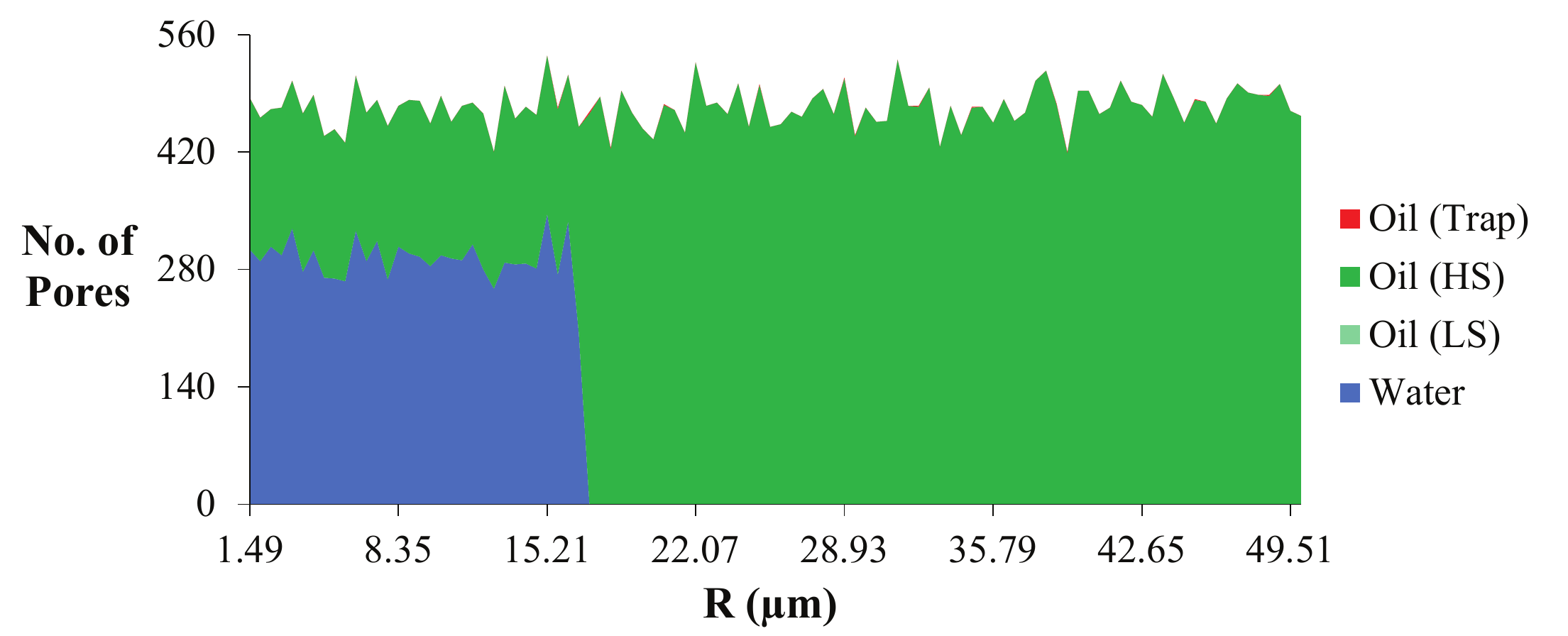}}}
\subfloat[\label{PBFOFigb}]{{\includegraphics[width=8cm]{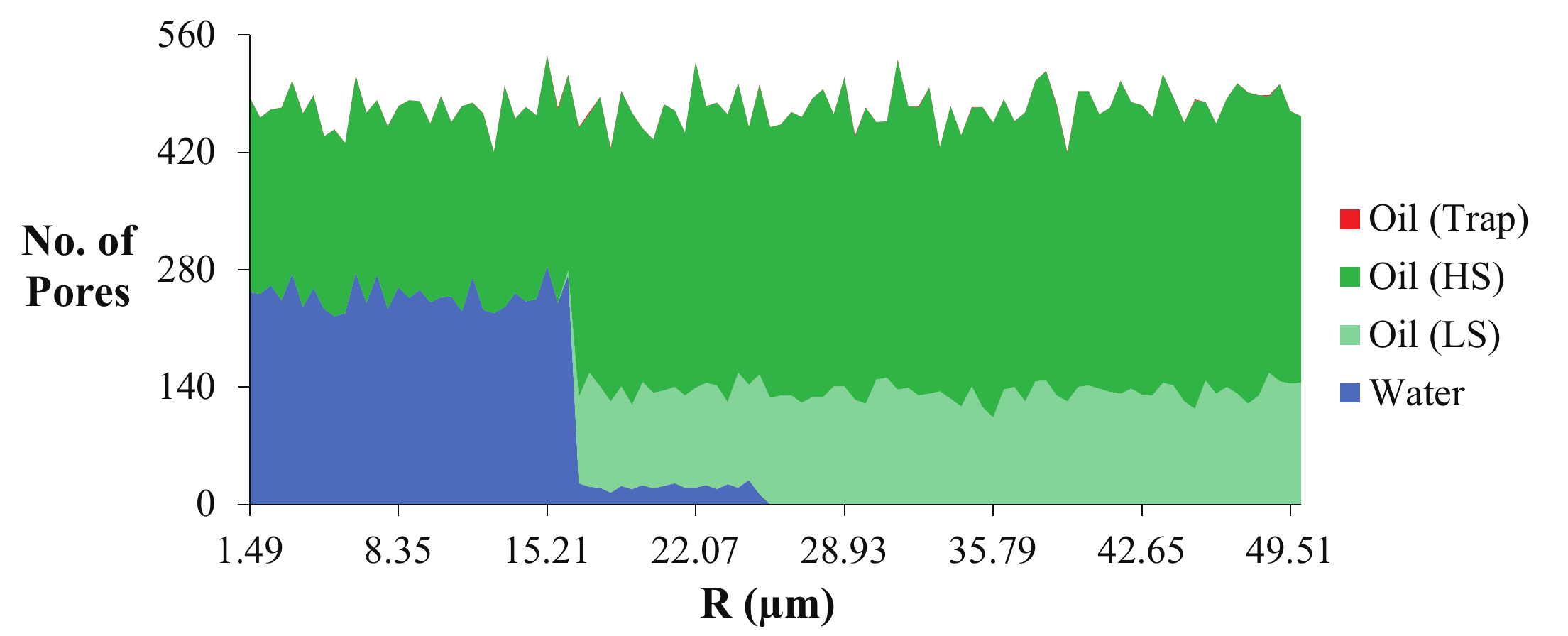}}}
\\
\subfloat[\label{PBFOFigc}]{{\includegraphics[width=8cm]{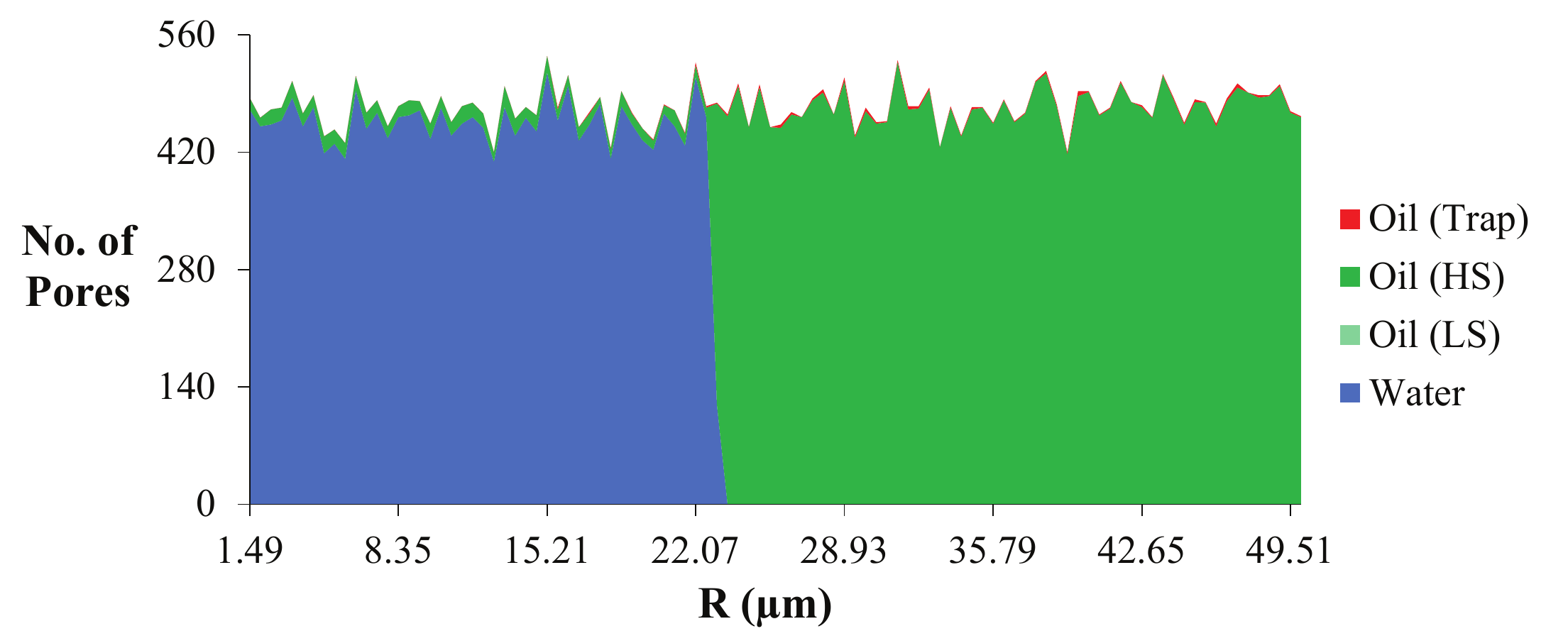}}}
\subfloat[\label{PBFOFigd}]{{\includegraphics[width=8cm]{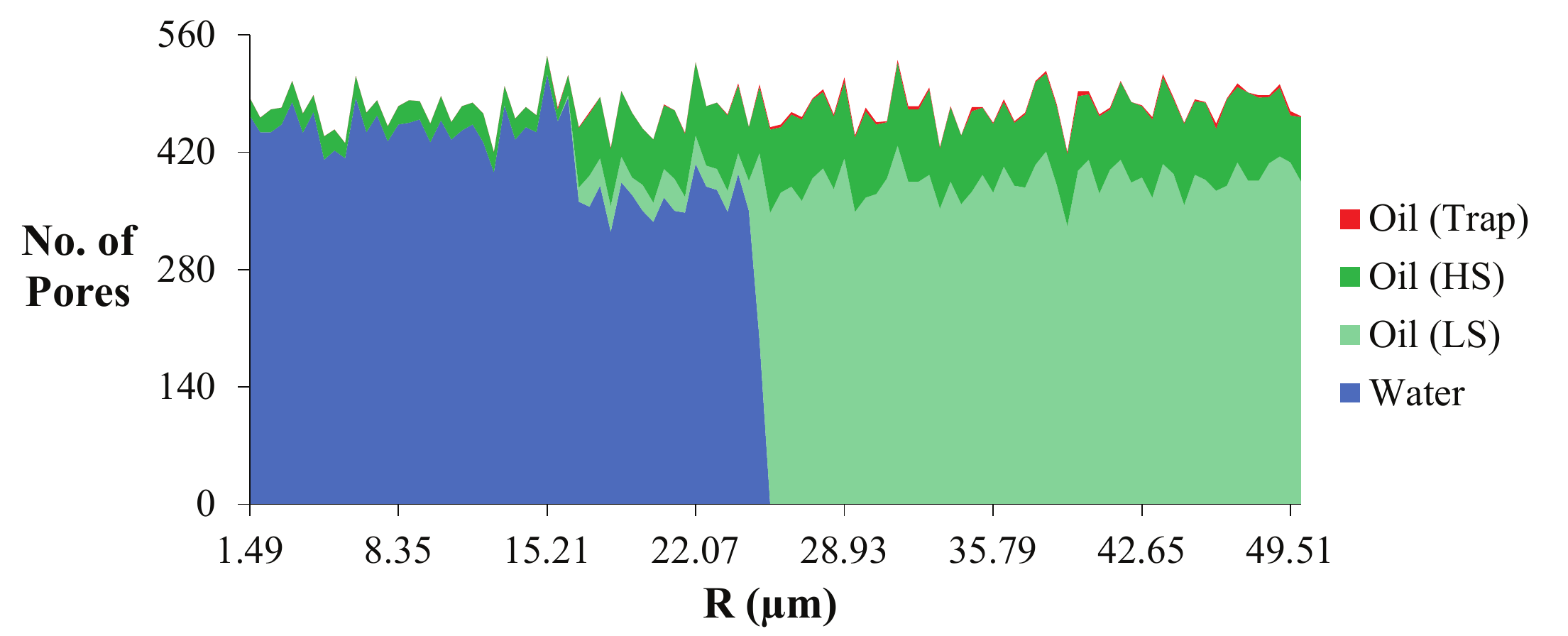}}}
\\
\subfloat[\label{PBFOFige}]{{\includegraphics[width=8cm]{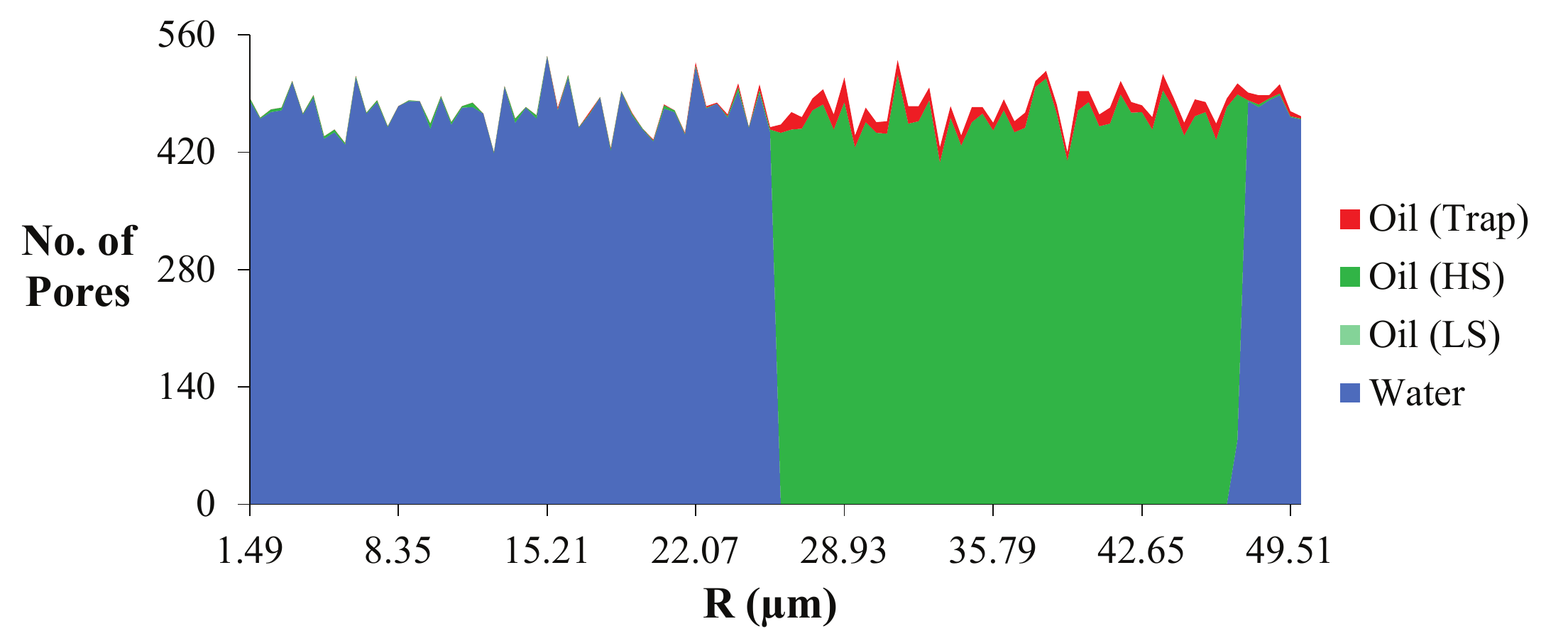}}}
\subfloat[\label{PBFOFigf}]{{\includegraphics[width=8cm]{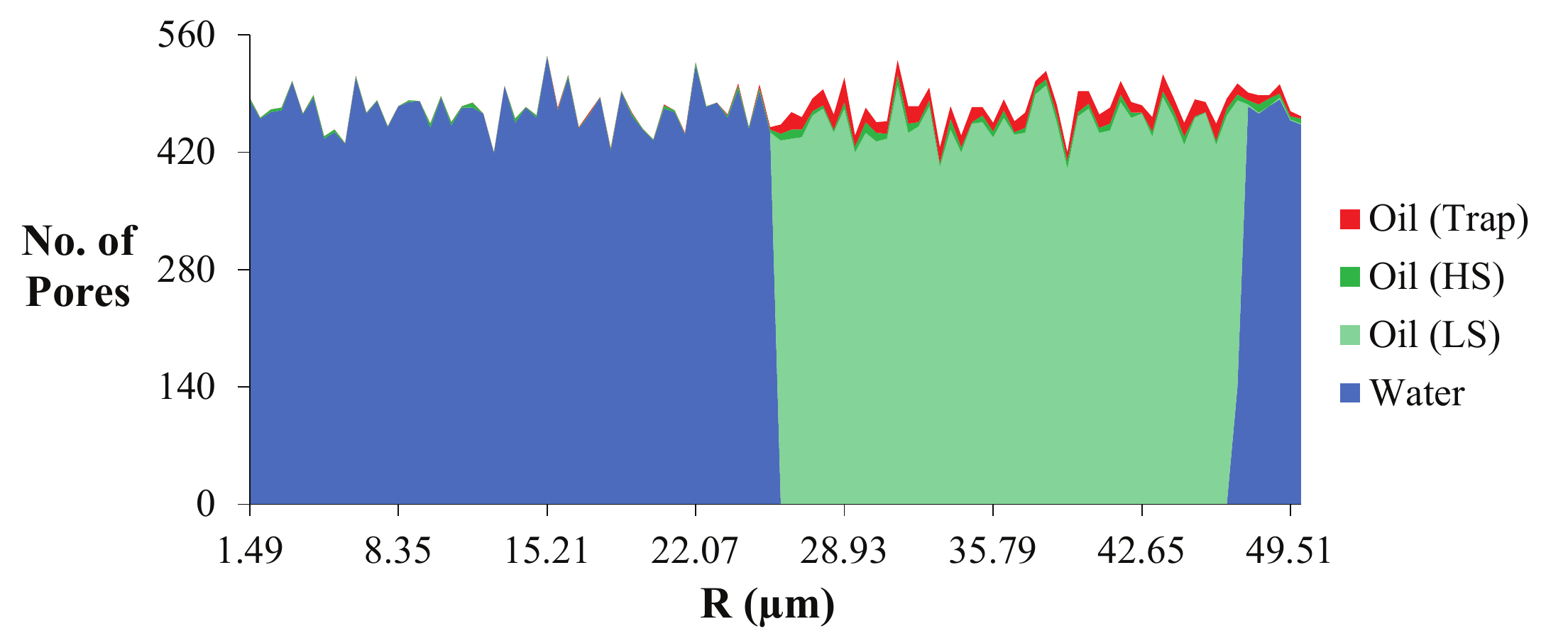}}}
\caption{Pore size fluid occupancy plots that correspond to the (a, c, e) HS brine injection and (b, d, f) LS brine injection $P_c$ curves presented in Figure~\ref{PBPcFig}. Plots show the fluid occupancy of pores arranged by capillary entry radius (a, b) at the first displacement step after water breakthrough, (c, d) at the time at which approximately 45\% of the pores contain water, and (e, f) at the time at which approximately 55\% of the pores contain water. The occupancies of pores are denoted by colours as follows: \textit{blue} water; \textit{dark green} oil (unmodified contact angle); \textit{light green} oil (modified contact angle); \textit{red} oil (trapped). \label{PBFOFig}}
\end{figure}

Fluid occupancies in pores of different sizes at various time points during the HS and LS simulations are presented in Figure~\ref{PBFOFig}, where pores are characterised as being either water-filled (blue), oil-filled with unmodified contact angle (dark green), oil-filled with modified contact angle (light green) or oil-filled and trapped (red). Recall that in a MWL system, the smallest pores are deemed to be WW (since $R_{min} =$ \SI{1}{\micro\metre} and $R_{max} =$ \SI{50}{\micro\metre}, the switch from WW to OW occurs at a pore radius of \SI{25.5}{\micro\metre}). During HS water imbibition, WW pores will generally be filled in order from smallest to largest according to the displacement rules defined in Section~\ref{SSDisp}; however, in the case where LS injection leads to dynamic contact angle modification, this pore filling sequence can be drastically altered by the associated changes in capillary entry pressures. This phenomenon can be clearly seen by comparing Figures~\ref{PBFOFiga} and \ref{PBFOFigb}, which show fluid occupancies shortly after breakthrough in the respective HS and LS simulations. While in the HS case only pores of radius up to approximately \SI{17}{\micro\metre} have been displaced, LS injection has recovered oil from WW pores of all sizes. This phenomenon, whereby the accessible WW pore radius is increased by reducing pore contact angles, is familiar from our previous work \citep{Wats17}, and has been shown to be governed by the following equation:
\begin{equation} 
	R_{LS,WWmax} = \frac{\cos\left(\theta_{LS,WW}\right)}{\cos\left(\theta_{HS,WW}\right)}R_{HS,WWmax}, \label{MagicEq}
\end{equation}
where $0 \leqslant \theta_{LS,WW} < \theta_{HS,WW} < \pi/2$. For a fixed $P_c$ value, $R_{LS,WWmax}$ and $R_{HS,WWmax}$ denote the radii of the largest WW pores that can be displaced in the presence and absence of contact angle modification, respectively. The ratio of the cosines is approximately 1.53 in this case, so $R_{LS,WWmax} \approx$ \SI{26}{\micro\metre} and it becomes clear why WW pores of all possible sizes have been displaced. This simple analysis reveals the reason why much of the oil can be displaced at a higher $P_c$ than previously possible, and also explains the period of recovery at fixed $P_c$ that is observed during LS brine imbibition. Since all WW pores that undergo contact angle modification \textit{immediately} satisfy their capillary entry condition, the gradual ingress of LS water after breakthrough leads to self-reinforcement of the imbibition displacement. Hence, without the requirement for any further reduction in $P_c$, a sizeable fraction of modified WW pores of all sizes is invaded (as well as a fraction of unmodified WW pores with radii up to \SI{17}{\micro\metre}). Note that while some OW pores will also have their contact angles reduced during the imbibition cycle, they remain inaccessible to the invading brine until the $P_c$ becomes sufficiently negative.

Following this period of equilibration in the LS simulation, further reduction in the $P_c$ is required to displace the remaining WW pores. A plot of the pore size fluid occupancy towards the end of the imbibition cycle is given in Figure~\ref{PBFOFigd}, with Figure~\ref{PBFOFigc} showing the HS simulation results at an equivalent stage. These figures indicate that, despite the significantly different pore filling sequences up to this point, the simulations have converged towards the same outcome --- that is, the majority of the WW pores in the network have been displaced. This highlights the important point that in heterogeneously-wetted networks of this type --- where LS injection does not alter the overall proportion of WW and OW pores --- the potential for additional oil recovery from the WW pores is likely to be very limited. Hence, any significant increment in oil production following LS injection will have to be obtained from the OW pores. Given that displacement of oil from OW pores by HS brine injection generally occurs in a volumetrically favourable largest to smallest sequence, this implies that any possible improvement in oil recovery by LS-induced wettability modification could only be achieved by increasing the fraction of OW pores that are displaced. Such an outcome would require a reduction in the topological trapping of oil within the OW pores of the network by a mechanism that we have termed the \textit{``microscopic sweep efficiency effect''} \citep{Wats17}.

Taking these factors into consideration, the question that remains is: why was the microscopic sweep efficiency of the OW pores not improved in this case? The answer lies in closer inspection of Figure~\ref{PBFOFigd}, which reveals widespread contact angle modification in the OW pores prior to the commencement of drainage in the LS simulation. Although these pores have become less strongly OW than those in the HS simulation (and are therefore displaced at a less negative $P_c$), in both the HS and LS drainage cycles the water is effectively invading a uniformly-wetted OW cluster and, consequently, the pore filling sequences are more-or-less identical (Figures~\ref{PBFOFige} and \ref{PBFOFigf}). Hence, an increase in microscopic sweep efficiency requires the capillary entry pressures of a sizeable fraction of the OW pores to be increased \textit{during} the drainage process. As remarked previously, this could potentially lead to \textit{immediate} displacement of these pores and a concurrent reduction in topological oil trapping as the LS front progresses across the network.

In our previous study of LS injection in uniformly wetted networks, we reported cases similar to that described above where, once again, no additional oil was recovered following dynamic contact angle reduction. In those cases, it was observed that incremental oil \textit{could} potentially be produced by narrowing the PSD of the network (i.e.\ reducing the value of $R_{max}/R_{min}$), increasing the extent of contact angle modification and/or shifting the initial network wettability towards a more neutral state. However, given the sequencing of pore-level displacements outlined above for the failure of LS brine injection to improve oil recovery, it is clear that none of these parameter modifications would significantly alter the outcomes reported for this particular scenario. Furthermore, it should be emphasised that the above results are not unique to the choice of a MWL system, and equivalent simulations for FW and MWS networks also demonstrate little or no additional oil recovery (results not shown). This is despite a significant increase in the HS brine saturation in the network in both the FW and MWS simulations prior to LS brine injection (7.0\% for FW and 12.8\% for MWS vs. 1.1\% for MWL). This suggests that by injecting LS brine only after HS brine has broken through, the HS brine in the network has limited capacity to suppress subsequent LS-induced wettability modification.

One method that could be used to improve the overall oil recovery by LS brine injection in the above cases would be to introduce the LS brine at a \textit{later} stage of the flood (i.e.\ \textit{during} the drainage phase). This approach would prevent the widespread modification of contact angles in the OW pores at an early stage of the displacement, and potentially allow for additional oil displacement by improving the microscopic sweep efficiency in the OW pores. We have performed simulations that support this theory (results not shown), but, in practice, the window of time available for such an approach to be successful may be narrow and, moreover, difficult to predict in an experimental setting. Indeed, other than investigations of tertiary LS brine injection following HS brine injection to irreducible oil saturation, we are not aware of any existing experimental studies that have attempted such an approach.

In \citet{Wats17}, our \textit{in silico} LS waterflooding approach demonstrated clear potential for a strong LSE in uniformly wetted pore networks. In contrast, the above investigation suggests that the potential for a LSE in non-uniformly wetted networks may be relatively limited. It is tempting to conclude that uniformly wetted networks may generally be better candidates for LS injection than non-uniformly wetted networks, but such a conclusion does not appear to be supported by the available experimental evidence. The coreflooding study of \citet{Ashr10}, for example, demonstrated significantly elevated oil recovery following secondary LS injection across a range of different initial core wettabilities. Furthermore, in a comprehensive review of the experimental LS coreflooding literature, \citet{Hamo15} noted that secondary LS injection produces more oil than secondary HS injection in the overwhelming majority of published studies. This evidence suggests that improved oil recovery by LS brine injection should be possible across the full spectrum of initial wettability states and indicates that our \textit{in silico} approach needs to be adapted to simulate a truly secondary LS injection protocol. These modifications are discussed next.

\subsection{Secondary Low Salinity Waterflooding \label{SecLSW}}
In the simulations performed in Section~\ref{PBLSW}, we assumed zero initial HS brine saturation ($S_{wi} = 0$) in our networks and injected HS brine until water breakthrough. This established a HS brine saturation in the network prior to the injection of LS brine. For secondary LS injection, however, HS brine is already present as connate water before the oil displacement commences. To create an initial HS brine saturation in our networks, we perform a steady-state primary drainage simulation that involves the injection of oil into a 100\% WW and 100\% water-filled network until the desired $S_{wi}$ is achieved. Note that this process follows the drainage methodology detailed in Section~\ref{SSDisp}, except that oil is now the invading phase. The initial HS brine will therefore be distributed uniformly throughout the network and reside predominantly in the smallest pores. We assign a non-uniform wettability state to the network only after the final $S_{wi}$ has been established (note that the pores that contain initial HS water are assumed to remain WW, even if we impose a FW or MWS wettability class).

Results from our earlier simulations of post-breakthrough LS injection suggest that incremental oil production will only be observed from non-uniformly wetted networks if contact angle modification can be sufficiently delayed to allow for a more efficient displacement of the OW pores. One candidate for the source of such a delay in secondary LS injection simulations could be the resident HS brine that will mix with the injected LS brine and inhibit localised changes in wettability. For this mechanism to be effective, it is important that the network is initialised with a realistic initial water saturation $S_{wi}$. To achieve this in our simplified networks, we narrow the width of the PSD and modify the volumetric scaling properties of the pores (through the volumetric exponent $\nu$). Additionally, we reduce the co-ordination number of the network $\bar{Z}$ to better reflect measured rock data --- this change also means that a significant fraction of pores can be filled with initial HS brine without approaching the percolation threshold of the network too closely (for 3D cubic systems, the percolation threshold is given by $\frac{3}{2\bar{Z}}$). If the fraction of initial HS water-filled pores were to be close to the percolation threshold, then breakthrough would occur shortly after the commencement of LS injection and the simulation protocol would, in effect, be the same as that used in Section~\ref{PBLSW}. We therefore make the following specific amendments to our base case parameter values in order to better mimic a more realistic pore system without having to incorporate additional layers of complexity associated with irregular network topologies and pore geometries at this stage (all other parameter values remain unchanged):
\begin{itemize}[label=--]
\item uniform PSD with minimum radius $R_{min} =$  \SI{1}{\micro\metre} and maximum radius $R_{max} =$  \SI{16}{\micro\metre};
\item we assume $V\left(R\right) \propto R^{0.5}$ to reflect a more physical scaling relationship for pore saturation values (note that we adopt a constant pre-factor to keep the total pore volume of the network consistent with the case where $\nu = 2$);
\item $\bar{Z} = 3.5$ (giving a percolation threshold of approximately 0.43, compared to 0.3 for $\bar{Z} = 5$);
\item $S_{wi} = 0.12$ (corresponding to HS brine in approximately 20\% of the pores).
\end{itemize}
Note that, in comparison to the networks used in Section~\ref{PBLSW}, all of the above modifications bring the features of the derived \textit{in silico} pore networks closer to those associated with real rock structures utilised in experimental coreflooding studies. We reiterate that we are not striving for quantitative prediction at this stage of the modelling: we are simply aiming to provide some insight into the underlying displacement sequences operating during LS floods. More intricate pore systems could be examined but inclusion at this juncture would complicate the interpretation of results without adding significant benefit.

\begin{figure}
\includegraphics[width=8cm]{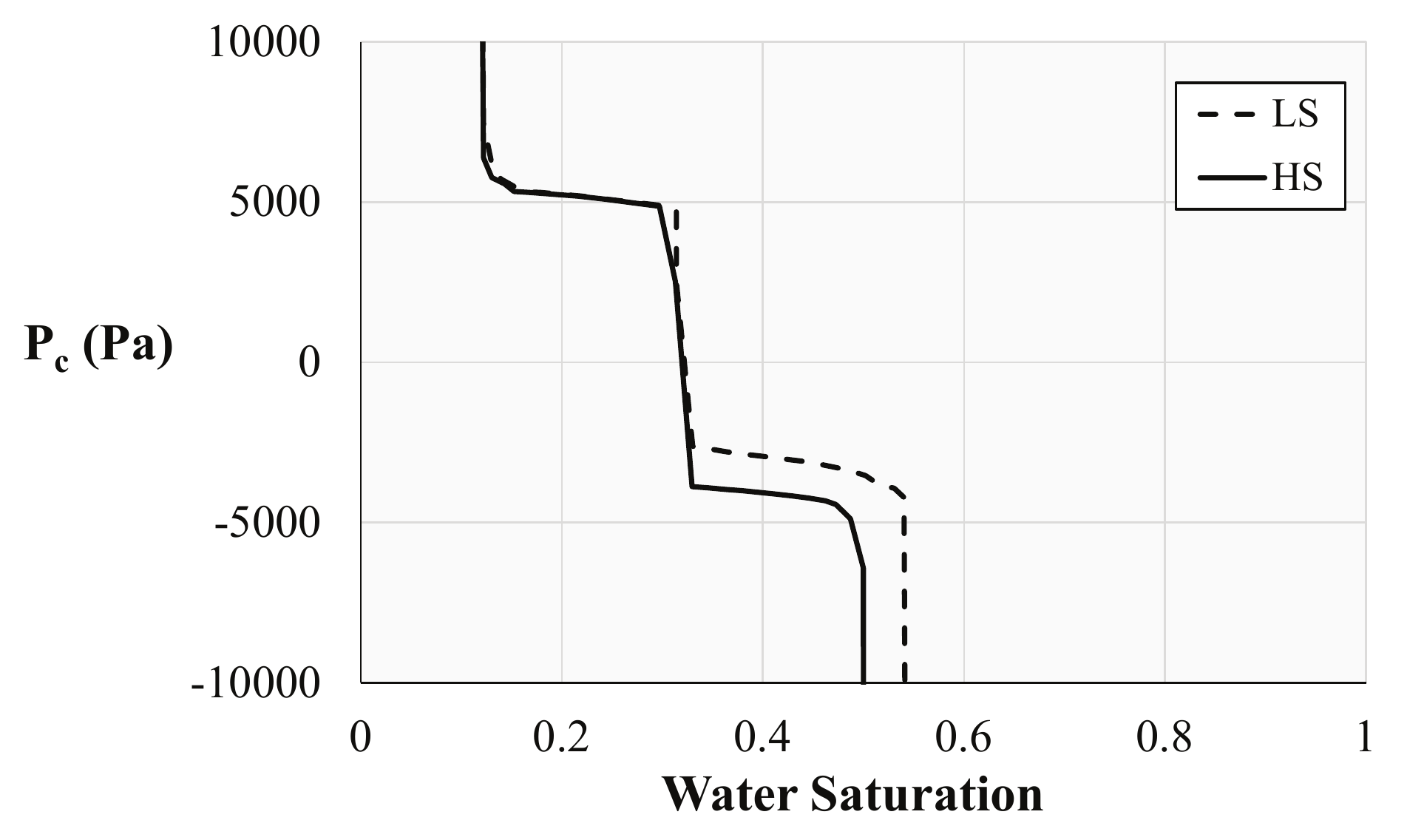}
\caption{Simulated $P_c$ curves for HS brine injection (solid line) and secondary LS brine injection (dashed line) into a MWL network with 50\% WW pores ($\theta_{HS,WW} = 60\degree$) and 50\% OW pores ($\theta_{HS,OW} = 140\degree$). \label{SecPcFig}}
\end{figure}

For consistency with the simulations in Section~\ref{PBLSW}, we focus on networks with the MWL wettability class (although note, again, that this is of little consequence to the qualitative outcomes of the study). The base case simulation again begins with 50\% WW pores ($\theta_{HS,WW} = 60\degree$) and 50\% OW pores ($\theta_{HS,OW} = 140\degree$), and ingress of LS brine again causes dynamic, localised reduction of contact angles ($\Delta \theta = 20\degree$). The $P_c$ curves from the HS and secondary LS brine injection simulations are presented in Figure~\ref{SecPcFig}, and it is immediately apparent that the qualitative features of the results are quite different to those observed previously (c.f.\ Figure~\ref{PBPcFig}). In contrast to the case where LS was introduced after breakthrough, here we see a limited response to LS injection throughout the imbibition phase and, moreover, an overall increase of approximately 5\% in the total volume of oil produced. This is a surprising result --- our earlier (post-breakthrough LS) simulations suggested that \textit{delayed} LS brine injection would be necessary to improve oil recovery but here (in secondary LS mode) additional oil has been obtained by injecting LS brine at an \textit{earlier} stage.

\begin{figure}
\subfloat[\label{SecFOFiga}]{{\includegraphics[width=8cm]{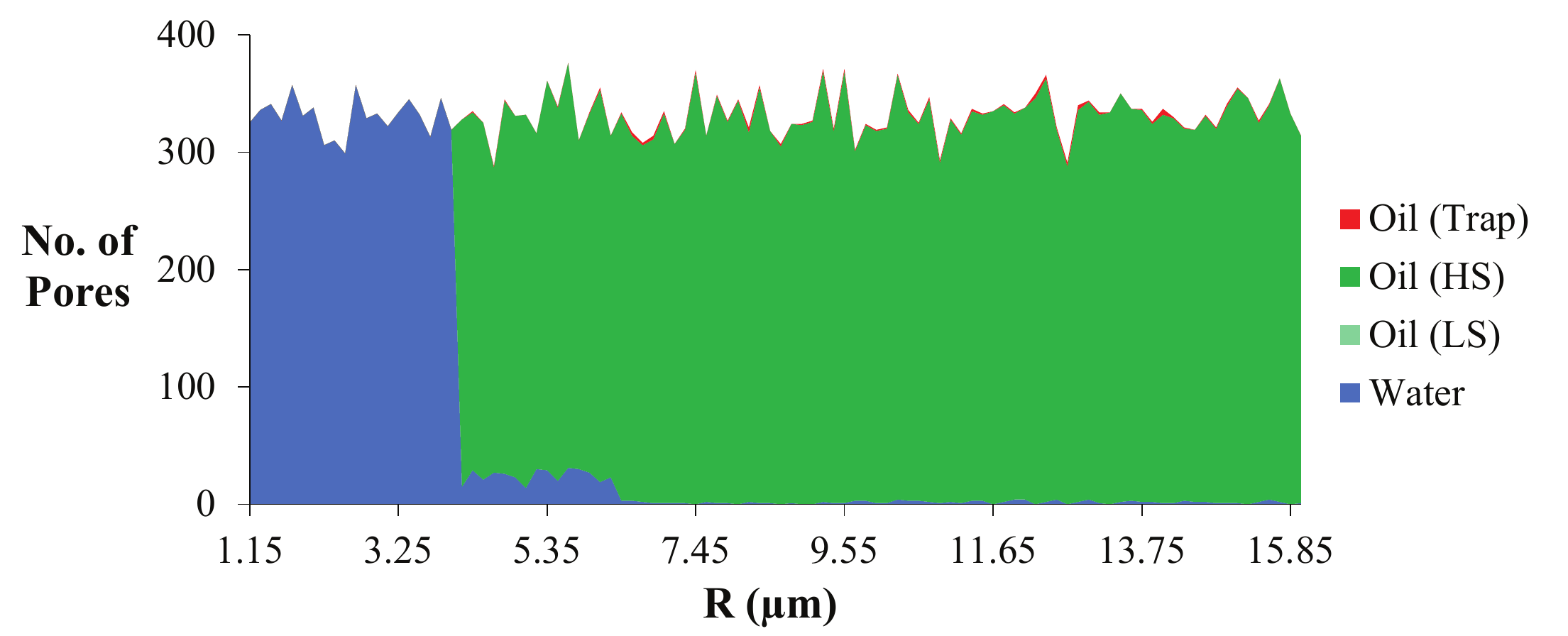}}}
\subfloat[\label{SecFOFigb}]{{\includegraphics[width=8cm]{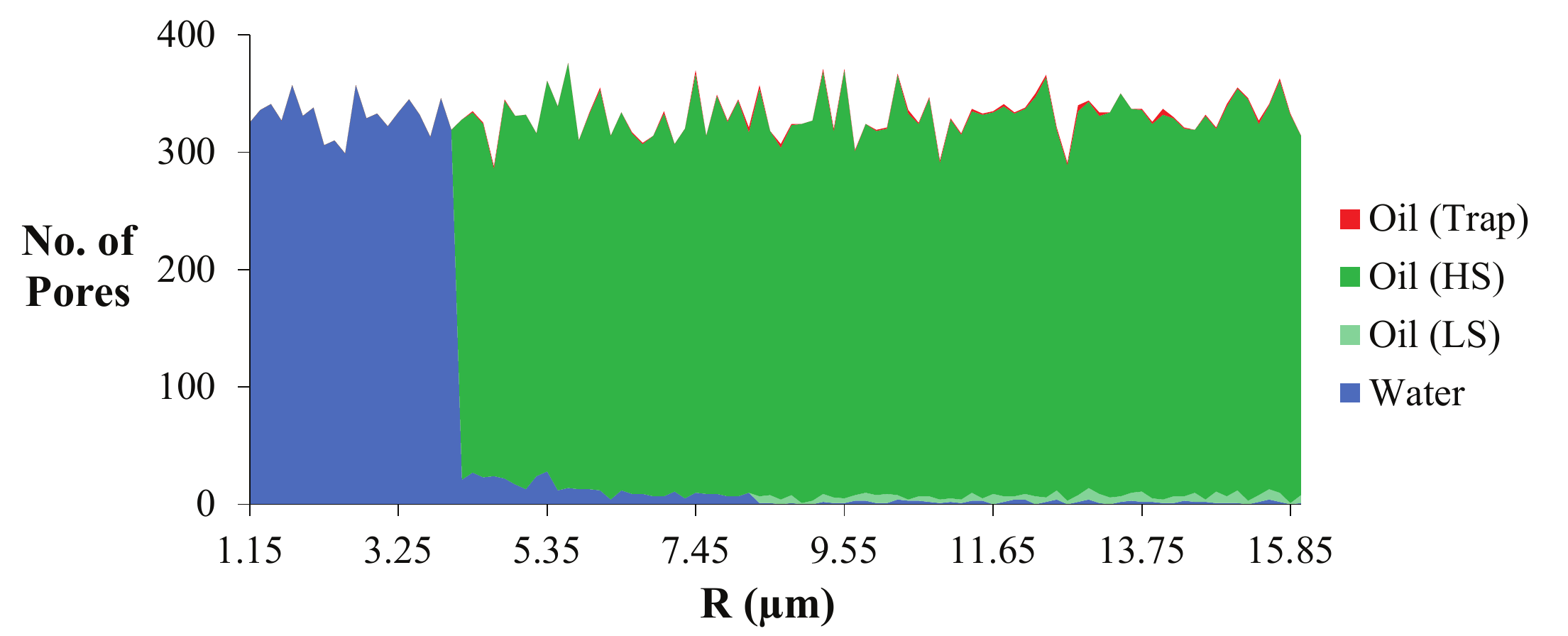}}}
\\
\subfloat[\label{SecFOFigc}]{{\includegraphics[width=8cm]{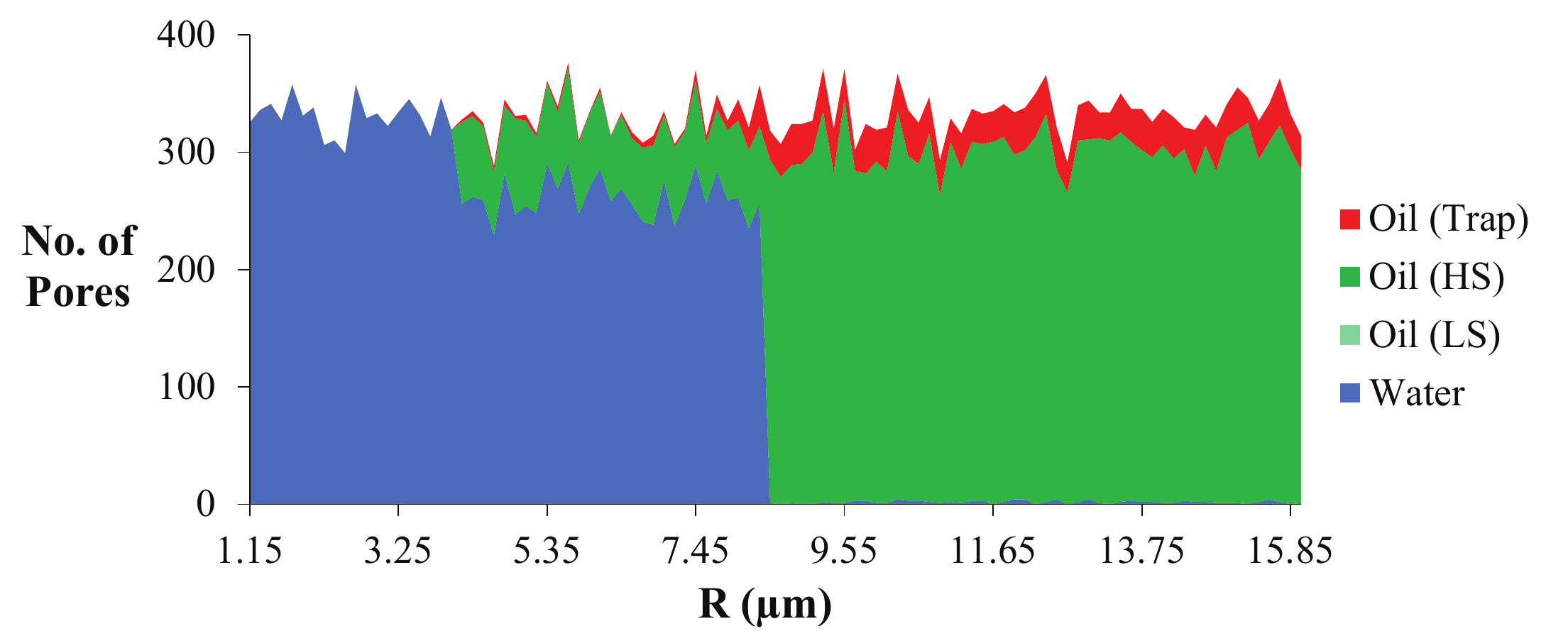}}}
\subfloat[\label{SecFOFigd}]{{\includegraphics[width=8cm]{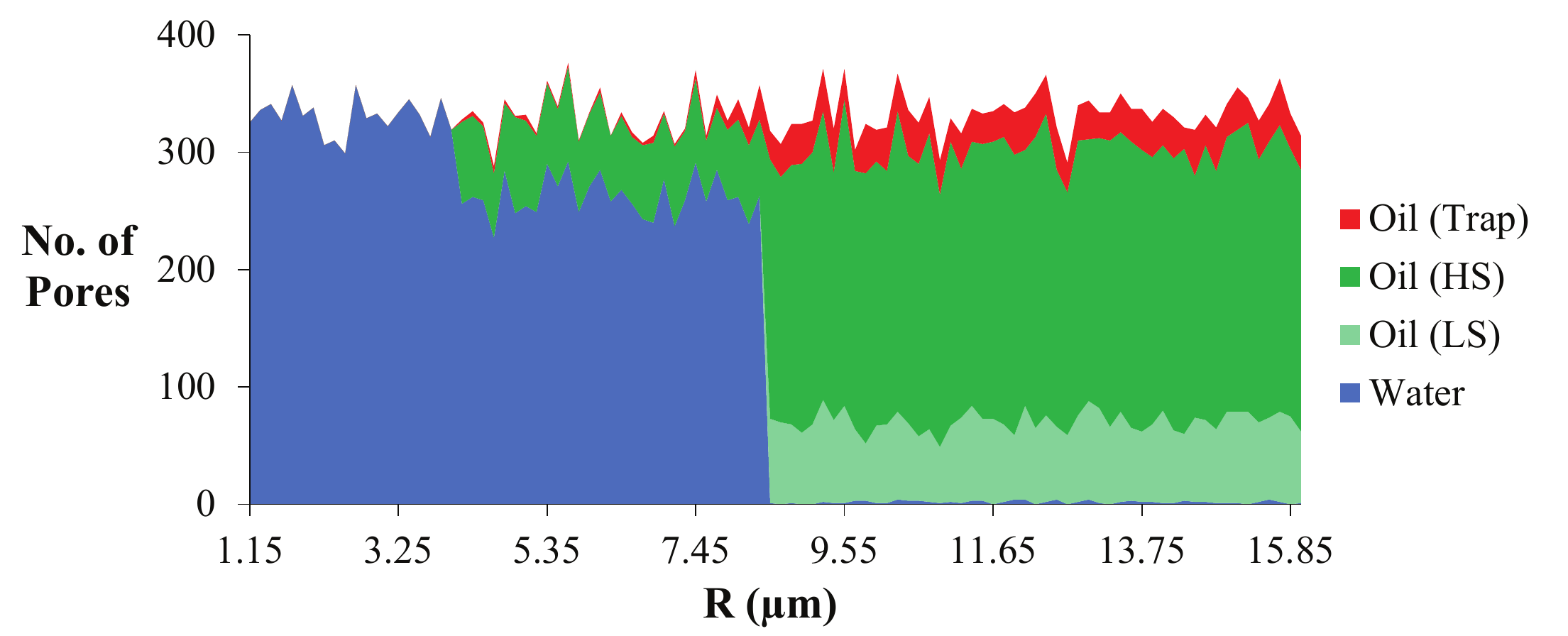}}}
\\
\subfloat[\label{SecFOFige}]{{\includegraphics[width=8cm]{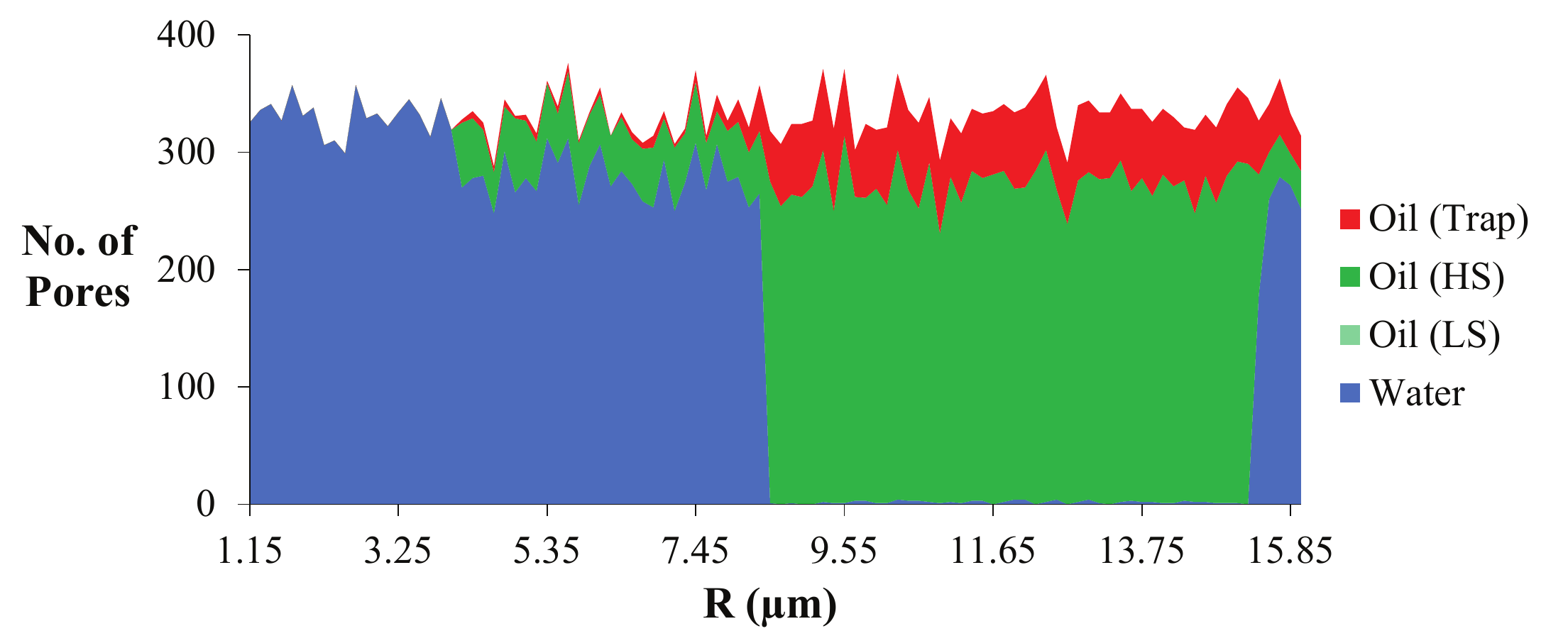}}}
\subfloat[\label{SecFOFigf}]{{\includegraphics[width=8cm]{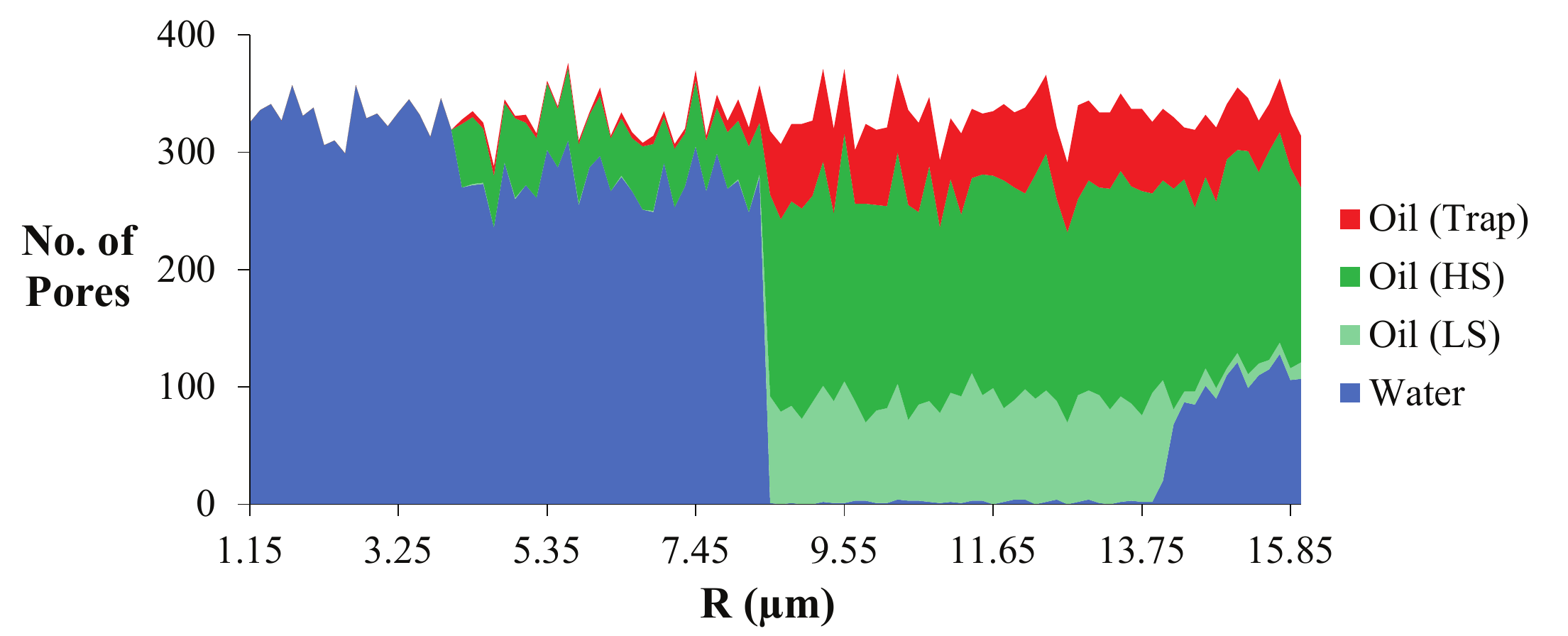}}}
\caption{Pore size fluid occupancy plots that correspond to the (a, c, e) HS brine injection and (b, d, f) LS brine injection $P_c$ curves presented in Figure~\ref{SecPcFig}. Plots show the fluid occupancy of pores arranged by capillary entry radius (a, b) at the first displacement step, (c, d) at the end of the imbibition cycle, and (e, f) at the time at which approximately 50\% of the pores contain water. The occupancies of pores are denoted by colours as follows: \textit{blue} water; \textit{dark green} oil (unmodified contact angle); \textit{light green} oil (modified contact angle); \textit{red} oil (trapped). \label{SecFOFig}}
\end{figure}

An examination of the pore filling sequences during the HS and LS displacements again allows us to understand the mechanisms responsible for this result. Pore size fluid occupancies at various stages of the HS and LS simulations are presented in Figure~\ref{SecFOFig}. The plots in Figure~\ref{SecFOFiga} (HS) and Figure~\ref{SecFOFigb} (LS) are taken at an early stage of the respective displacements and highlight the fact that secondary LS brine injection causes the pore filling sequences to diverge from the very outset. At this stage, some of the \textit{largest} WW pores (which are proximal to the network inlet) have already been invaded by LS due to the reduced connectivity of the pore space. Recall, however, that the pores displaced over the entire imbibition cycle should be effectively identical for both HS and LS brine injection. Figure~\ref{SecFOFigc} (HS) and Figure~\ref{SecFOFigd} (LS), which are plotted at the end of the respective imbibition cycles, prove this to be the case. The more important issue at this stage is the extent of contact angle modification in the OW pores, and a comparison of Figures~\ref{PBFOFigd} and \ref{SecFOFigd} reveals that, in the case of secondary injection, significantly more OW pores remain unmodified by LS brine as drainage begins. At this stage of the secondary LS injection simulation, breakthrough has already occurred and a reasonable amount of LS water (0.33 pore volumes) has been injected into the system. Hence, the lack of contact angle modification can only be explained by the ongoing influence of the connate HS brine in the system coupled with the reduced connectivity.

\begin{figure}
\subfloat[\label{SecFOEndFiga}]{{\includegraphics[width=8cm]{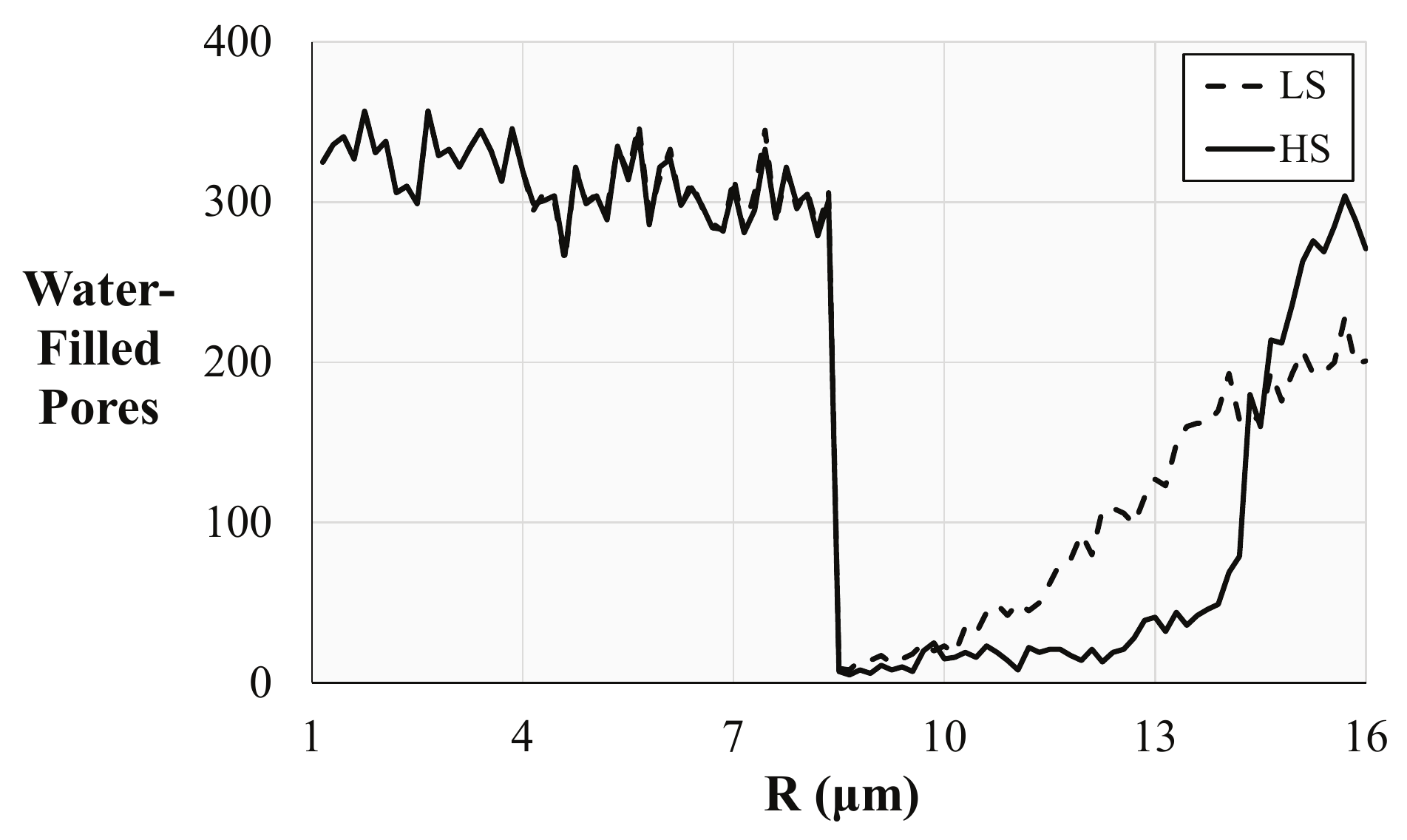}}}
\quad
\subfloat[\label{SecFOEndFigb}]{{\includegraphics[width=8cm]{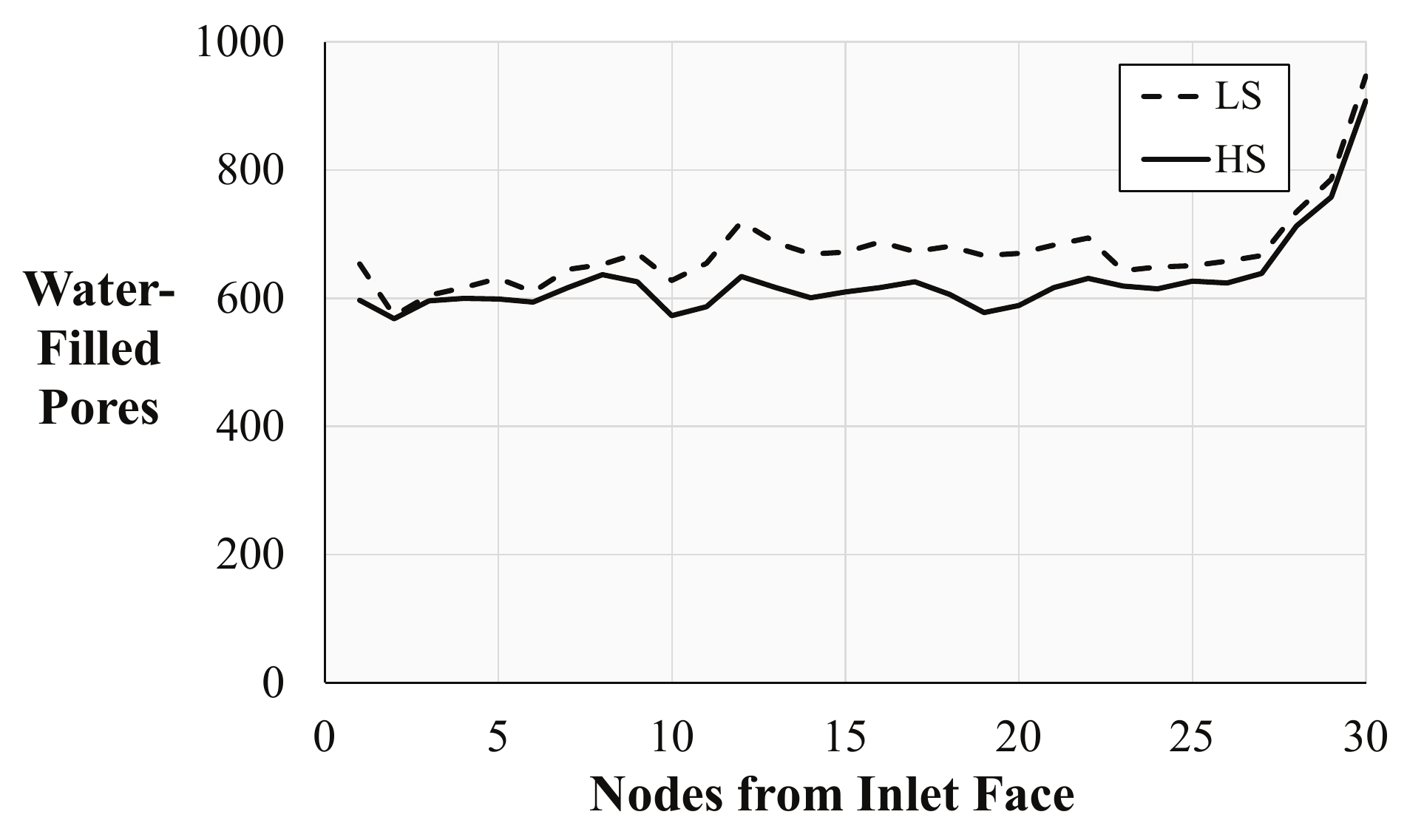}}}
\caption{Plots of (a) the size distribution and (b) the spatial distribution of water-filled pores at residual oil saturation for simulations of HS brine injection (solid line) and secondary LS brine injection (dashed line) into a MWL network with 50\% WW pores ($\theta_{HS,WW} = 60\degree$) and 50\% OW pores ($\theta_{HS,OW} = 140\degree$). \label{SecFOEndFig}}
\end{figure}

As highlighted previously, a delay in LS-induced contact angle modification in the OW pores of a non-uniformly wetted network can be beneficial for oil recovery because it allows a more efficient displacement of the remaining oil to be achieved. The mechanism of improved microscopic sweep efficiency can be seen in Figure~\ref{SecFOFigf}, where the drainage cycle in the LS simulation proceeds by filling successively smaller pores with modified contact angles (unmodified pores initially remain inaccessible). Gradual reduction of the $P_c$ allows this to continue and, as more and more pores are contacted by LS, any modified pore with radius larger than the smallest accessible pore will automatically satisfy its capillary entry threshold and be immediately invaded. In volumetric terms, this sequence of pore filling is less favourable than the equivalent HS simulation (Figure~\ref{SecFOFige}); however, the LS-induced phenomenon of immediate invasion increases the total fraction of pores displaced (62.2\% for LS vs. 58.1\% for HS) and ultimately increases the overall oil recovery. Figure~\ref{SecFOEndFig} displays the final distributions of water-filled pores for the HS and LS simulations organised by both pore entry radius (Figure~\ref{SecFOEndFiga}) and by spatial position in the network (Figure~\ref{SecFOEndFigb}). Figure~\ref{SecFOEndFiga} shows that, despite increased trapping of oil pores with radii in the range 14--16 \si{\micro\metre}, the LS simulation has improved overall oil recovery by significantly increasing the displacement of oil-filled pores with radii in the range 10--14 \si{\micro\metre}. Figure~\ref{SecFOEndFigb}, meanwhile, confirms that this increase in microscopic sweep efficiency is not a localised effect. Compared to the injection of HS brine, LS brine injection has increased the displacement of oil-filled pores across the entire width of the network.

Our results in Section~\ref{PBLSW} suggested that a significant improvement in oil recovery in non-uniformly wetted networks would require an explicit delay in the injection of LS brine, but the above simulation indicates that significant incremental oil can be produced even if LS brine is injected in secondary mode. This may appear to be something of a contradiction but recall that we have made quite different assumptions in these two cases. The distribution of HS brine in the network prior to LS injection is one aspect that has changed considerably. For post-breakthrough LS injection, the injected HS brine spans the network and is largely distributed in inlet-connected clusters, whereas for secondary LS injection the connate HS brine is both non-spanning and widely dispersed throughout the pore structure. The secondary LS base case simulation suggests that the influence of the connate HS brine plays a critical role in the success of secondary LS injection, and we investigate this below by performing sensitivity simulations on the initial HS brine saturation $S_{wi}$.

\begin{figure}
\includegraphics[width=8cm]{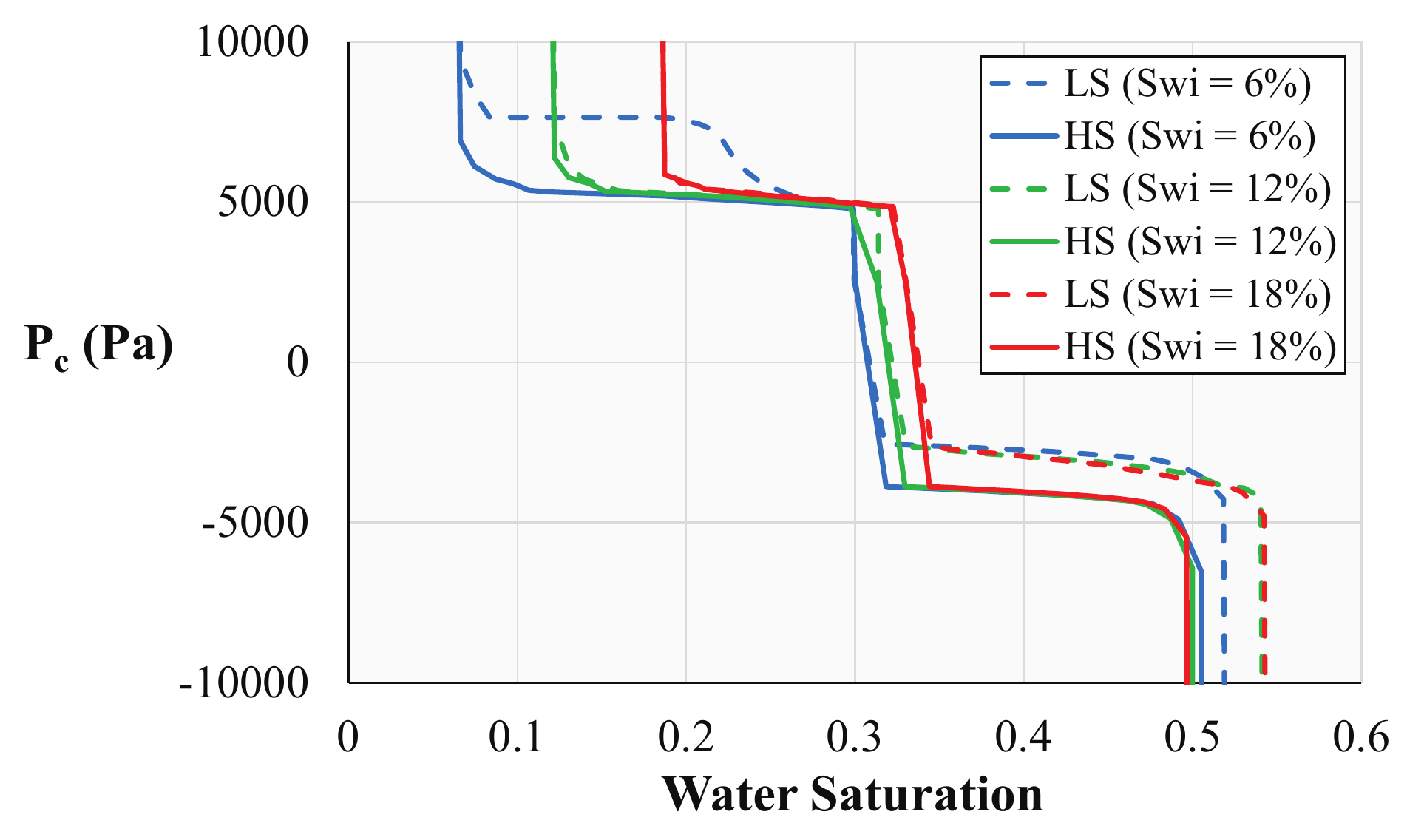}
\caption{Simulated $P_c$ curves for HS brine injection (solid lines) and secondary LS brine injection (dashed lines) into a MWL network initialised with a range of $S_{wi}$ values. Results for the base case parameter value ($S_{wi} =$ 12\%; blue lines) are compared to results for both a smaller ($S_{wi} =$ 6\%; green lines) and a larger ($S_{wi} =$ 18\%; red lines) value of $S_{wi}$. The network contains 50\% WW pores ($\theta_{HS,WW} = 60\degree$) and 50\% OW pores ($\theta_{HS,OW} = 140\degree$). \label{SecSwiPcFig}}
\end{figure}

We have studied the impact of both a decrease ($S_{wi} =$ 6\%; $\sim$10\% of available pores) and an increase ($S_{wi} =$ 18\%; $\sim$30\% of available pores) in the initial HS water saturation. $P_c$ curves from the HS and LS simulations are compared to those from the base case simulation in Figure~\ref{SecSwiPcFig}. We find that a reduction in $S_{wi}$ has a negative impact on the efficacy of oil recovery by LS brine injection. The profile of the $P_c$ curve during the imbibition phase indicates an accelerated rate of contact angle modification within the network compared to the LS base case simulation. This limits the scope for an efficient displacement of the OW pores, and ultimately results in a smaller increment in the total volume of oil produced (1.9\% vs.\ 4.9\% for the base case LS simulation). An increase in $S_{wi}$, on the other hand, could be expected to significantly delay modification of contact angles by the injected LS brine and ultimately lead to greater incremental oil production than the base case LS simulation. However, this does not prove to be the case and, in fact, the results from the two simulations are remarkably similar. This appears to be related to the fact that an increase in $S_{wi}$ leads to earlier water breakthrough (after 0.030 pore volumes water injection rather than 0.065 pore volumes water injection in the base case LS simulation): the elevated ingress of LS brine after breakthrough accelerates the removal of HS brine from the network and leads to a rate of LS-induced contact angle modification that is very similar to the base case simulation.

\begin{figure}
\includegraphics[width=8cm]{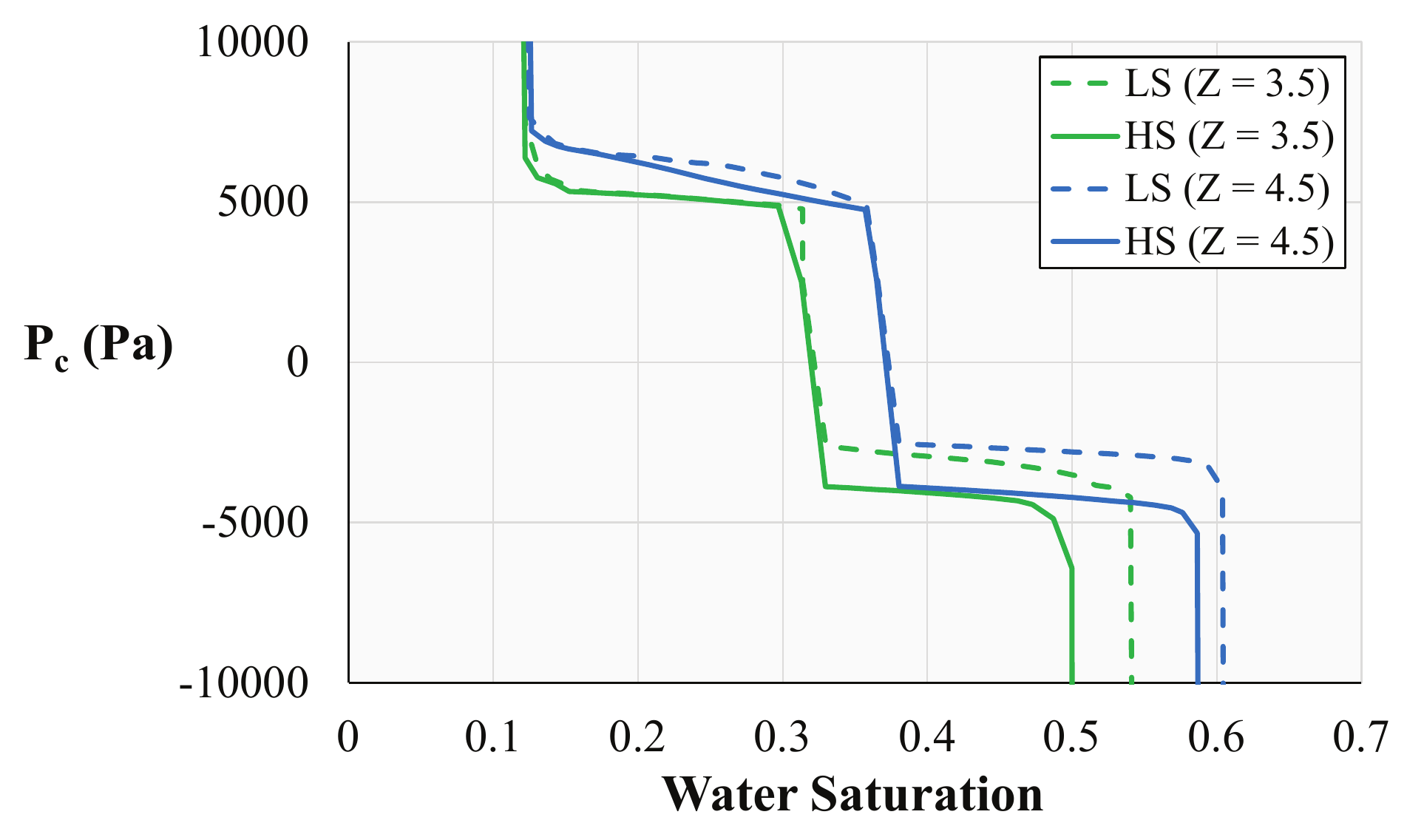}
\caption{Simulated $P_c$ curves for HS brine injection (solid lines) and secondary LS brine injection (dashed lines) into MWL networks with different $\bar{Z}$ values. Results for the base case parameter value ($\bar{Z} = 3.5$; green lines) are compared to results for $\bar{Z} = 4.5$ (blue lines). The networks both contain 50\% WW pores ($\theta_{HS,WW} = 60\degree$) and 50\% OW pores ($\theta_{HS,OW} = 140\degree$). \label{SecZPcFig}}
\end{figure}

This simulation with $S_{wi} =$ 18\% is particularly interesting because it demonstrates that LS brine injection can displace significant incremental oil from non-uniformly wetted networks, even when the LS brine is injected close to the point of water breakthrough. Our earlier simulations in Section~\ref{PBLSW} had suggested that this phenomenon was unlikely, so it appears that the changes we have made --- other than modifying the water injection protocol --- may also be influential in the positive LSE that we have observed in the above secondary LS injection simulations. It seems unlikely that narrowing the width of the PSD would produce such dramatic consequences for the efficacy of LS injection. Indeed, according to results from \citet{Wats17}, the potential for an efficient displacement of the OW pores should increase as the ratio of the largest OW pore radius to the smallest OW pore radius is reduced. For a change in the PSD from $U\left(1,50\right)$ to $U\left(1,16\right)$ in a MWL network with $\alpha = 0.5$, this ratio falls only slightly from 1.96 (i.e.\ \SI{50}{\micro\metre} / \SI{25.5}{\micro\metre}) to 1.88 (i.e.\ \SI{16}{\micro\metre} / \SI{8.5}{\micro\metre}). Therefore, we focus instead on the connectivity of the pore network and perform a sensitivity simulation on the co-ordination number $\bar{Z}$. Figure~\ref{SecZPcFig} compares the base case HS and LS $P_c$ curves with those from a simulation with increased network connectivity ($\bar{Z} = 4.5$). The simulations for $\bar{Z} = 4.5$ both recover more oil than their equivalent simulations for $\bar{Z} = 3.5$ because of reduced topological oil trapping. More importantly, however, the increment in the volume of oil displaced by LS brine injection for $\bar{Z} = 3.5$ is much greater than that for $\bar{Z} = 4.5$ (4.9\% vs.\ 1.6\%). This indicates that pore contact angles are less readily modified by LS brine in the more poorly connected network, which is likely due to a combination of delayed water breakthrough and increased residence time of HS water as it navigates circuitous pathways through the pore structure towards the outlet. Network connectivity can therefore have a crucial influence on the LSE in non-uniformly wetted networks, and the lack of incremental oil recovery in our post-breakthrough LS brine injection simulations can be largely explained by the well-connected network that was utilised for the study.

\section{Discussion \label{Disc}}
The mechanisms and efficacy of LS waterflooding have been topics of much recent debate in the oil industry. While the LSE is yet to be fully understood, a consensus has emerged that injected LS brine operates by altering the wettability of its surrounding rock structure. We have recently developed novel pore-scale models to investigate the dynamic LSE in uniformly wetted pore networks \citep{Wats17, Bouj18}, and we have extended our methodology here to study networks of non-uniform initial wettability. Our results indicate that the conditions required for a positive LSE in non-uniformly wetted networks are typically more stringent than those for uniformly wetted networks and, moreover, we have identified several potential reasons for the inconsistent performance that is observed in experimental LS waterflooding studies.

Following \citet{Wats17}, we have simulated LS waterflooding by coupling a steady-state oil displacement model to a tracer algorithm that estimates the spatio-temporal evolution of water salinity as HS and LS brines mix in the \textit{in silico} pore networks. We have chosen not to consider explicit chemical reactions in the COBR system and we have instead captured the apparent \textit{consequences} of LS brine injection by assuming that brine salinities beneath a critical threshold induce localised wettability modification. Changing the contact angles of individual pores in this way influences the sequence of pore filling by the invading LS brine and, under certain circumstances, increases the total volume of oil produced during the displacement.

While the steady-state LS waterflooding model that we have reported here is less sophisticated than the unsteady-state model of \citet{Bouj18}, the steady-state approach does have certain benefits. In particular, the steady-state model can simulate the displacement of fluids from large 3D pore networks at substantially lower computational cost than the unsteady-state model --- this is significant because a meaningful investigation of the LSE in regular non-uniformly wetted pore networks can \textit{only} be achieved using 3D simulations. For a regular pore network to contain spanning clusters of both WW and OW pores, the percolation threshold of the network must be $\leqslant 50\%$. However, the percolation threshold of a regular 2D pore network is always $\geqslant 50\%$. So, for regular 2D pore networks, it is therefore only possible to have spanning clusters of both WW and OW pores for the very restrictive case where $\alpha = 0.5$ (50\% WW and 50\% OW pores), the network is fully connected ($\bar{Z} = 4$) and the WW and OW pores are randomly distributed (irregular, poorly-connected 2D networks would be even more compromised). Moreover, were this restrictive case to be used to investigate LS waterflooding, model simulations would provide little or no insight into the mechanisms of the LSE. Indeed, for any LS waterflooding simulation where individual pores maintain their initial wetting \textit{class} (i.e.\ WW or OW), oil from most of the 50\% WW pores would be displaced during the imbibition cycle. Then, at the end of the imbibition cycle, both the fraction of water-filled pores and the fraction of oil-filled pores would be close to the network percolation threshold (50\%), meaning that most of the remaining oil would be isolated from the network outlet and trapped inside the OW pores. Hence, only oil from the WW pores could ever be displaced and there would be no scope for LS brine to improve oil recovery by increasing microscopic sweep efficiency in the OW pore fraction.

For consistency with our earlier work in uniformly wetted media, we began this study by simulating an \textit{early tertiary} LS waterflooding protocol where LS brine was introduced into the network after breakthrough of injected HS brine. We initialised a network with an equal balance of WW (60\degree) and OW (140\degree) pores and assumed a contact angle reduction of 20\degree\ in all oil-filled pores that were contacted by water of sufficient freshness (note the implication that no OW pores could become WW over the course of the simulation). The assumed extent of wettability modification by LS brine is consistent with the recent experimental findings of \citet{Khis17}, who observed an average LS-induced contact angle reduction of around 16\degree\ in miniature sandstone core samples. While this simulation ultimately produced the same volume of oil as the equivalent case with only HS brine injection, the results did provide valuable insight into the necessary conditions for improved recovery by LS brine injection in non-uniformly wetted networks of this type. The simulation demonstrated that any action of the LS brine on the oil-filled WW pores is essentially superfluous, since WW pores will always be displaced by imbibition regardless of their contact angle or the composition of the injected brine. Correspondingly, the OW pores contain the only viable source of additional oil and, due to the volumetrically-favourable nature of the standard OW pore filling sequence, any increase in the overall oil recovery by LS brine injection can only be achieved by improving the microscopic sweep efficiency in the OW pores.

The above requirement for improved microscopic sweep efficiency in the OW pores is particularly stringent, and it has several implications for oil recovery by LS brine injection in non-uniformly wetted networks. Firstly, increased microscopic sweep efficiency is a phenomenon that is associated only with dynamic, LS-induced contact angle \textit{reduction} \citep{Wats17}. Hence, if LS brine injection were to lead to an \textit{increase in oil-wetness} --- a phenomenon that has been reported on several occasions \citep{Buck97, Sand11} --- there would be \textit{no potential} for an associated improvement in oil recovery in a non-uniformly wetted network. Note that we refer here only to cases where the modified WW pores remain WW; cases where contact angles cross 90{\degree} are less clear-cut and will be discussed in detail below. Secondly, increased microscopic sweep efficiency in the OW pores requires a sizeable fraction of OW contact angles to be reduced \textit{during the drainage cycle}. Hence, parameters that contribute to the delay of widespread contact angle reduction by injected LS brine in non-uniformly wetted networks are crucial to the production of incremental oil. Note that we did not encounter this type of restriction in our earlier studies of LS brine injection in uniformly wetted networks, where parameters such as the PSD, the initial network wettability and the extent of contact angle modification were identified as the most critical parameters in determining the magnitude of any positive LSE. For non-uniformly wetted networks, these parameters can still have an important influence on the extent of any positive LSE. However, it is only the PSD and the contact angle properties of the \textit{OW pore fraction} that are significant in this regard, and even then only under the condition that wettability modification by injected LS brine can be sufficiently delayed.

The observation that the potential for improved oil recovery by LS injection requires a stricter set of conditions for non-uniformly wetted networks than for uniformly wetted networks prompted us to adopt a more experimentally-relevant LS brine injection protocol and to focus on pore networks with more physically realistic features. Consequently, in Section~\ref{SecLSW}, we adjusted the properties of the pore space (i.e.\ PSD, $\bar{Z}$, $\nu$) to ensure that a realistic HS $S_{wi}$ could be initialised in the network and we simulated \textit{secondary LS brine injection} rather than a post-breakthrough LS brine injection approach. These changes to our methodology and parameters allowed us to perform a more thorough investigation of the LSE in non-uniformly wetted pore networks. For parameter values that were otherwise identical to the earlier post-breakthrough LS brine injection simulation, we found that secondary LS brine injection led to a marked increase in oil recovery compared to HS brine injection alone. Although LS brine was injected \textit{earlier}, the modification of contact angles in many OW pores was \textit{delayed} and a more efficient displacement of the OW pore space was ultimately achieved. This result highlights that the parameters that we modified --- most notably $S_{wi}$ and $\bar{Z}$ --- are crucial factors that can control the extent of any observed LSE in non-uniformly wetted networks.

We explored the impact of HS $S_{wi}$ on the efficacy of secondary LS brine injection by performing sensitivity simulations for $S_{wi}$ values that were both smaller and larger than the base case value. For the non-uniformly wetted network that we considered, our results indicated an interesting nonlinear relationship between $S_{wi}$ and the volume of incremental oil recovered by LS brine injection. When the initial fraction of HS brine-filled pores was far from the network percolation threshold, the additional oil recovered by LS brine injection increased with increasing $S_{wi}$. This result reflects the fact that increasing the connate HS brine saturation increases the delay in LS-induced wettability modification and therefore improves the microscopic sweep efficiency during the drainage cycle. Interestingly, the trend of increasing oil recovery by secondary LS brine injection with increasing HS $S_{wi}$ is consistent with experimental results reported by \citet{Zhan06} for waterflooding of Berea sandstone cores with permeabilities of 500 mD and 1100 mD. As more pores in our networks were filled with initial water and the fraction of HS brine-filled pores approached the percolation threshold, the trend for increasing incremental recovery with increasing $S_{wi}$ saturated and our results showed a marginal decline in the efficacy of LS brine injection. This result appears to reflect the fact that early water breakthrough --- and the associated egress of HS brine from the network --- can accelerate the localised modification of wettability by injected LS brine and consequently reduce the potential for an efficient displacement of the OW pore fraction.

We also investigated the impact of the network co-ordination number $\bar{Z}$ on the efficacy of secondary LS brine injection, and our simulations showed an increase in the relative volume of incremental oil recovered by LS brine injection as $\bar{Z}$ was reduced. For the non-uniformly wetted networks that we considered here, this result reflects the fact that a reduction in the pore space connectivity increases the residence time of connate HS brine in the network and delays the widespread modification of wettability until the drainage stage of the waterflood. The trend of increasing efficacy of secondary LS brine injection with decreasing $\bar{Z}$ is not necessarily restricted to non-uniformly wetted networks, however, as unpublished simulation results indicate a similar result for uniformly wetted pore networks. This is because oil trapping is prevalent for poorly connected networks of all initial wettability states and, hence, the \textit{potential} for LS brine injection to improve microscopic sweep efficiency increases as $\bar{Z}$ is reduced. It would be interesting to test this theory against experimental coreflooding observations, but the average pore space connectivity is a parameter that is not typically considered in laboratory studies. The permeability of the core is generally reported, but neither the network connectivity nor the network PSD can be explicitly determined from this measurement. Our simulation results indicate that both $\bar{Z}$ and the distribution of pore sizes can influence the LSE by distinct pore-scale mechanisms and an ideal experimental analysis should therefore aim to control for these factors independently.

A potentially important parameter that we have yet to explicitly consider in this study is the fraction $\alpha$ of OW pores in our \textit{in silico} networks. Given that any incremental oil recovered by LS brine injection into non-uniformly wetted networks should be mostly recovered from the OW pore fraction, the value of $\alpha$ seems likely to be a key factor in determining the extent of any LSE (note that we refer only to values of $\alpha$ that give spanning clusters of both OW and WW pores; c.f.\ Figure~\ref{PercFig}). Based on the results reported in Section~\ref{Results}, our simulations would predict that the potential for improved oil recovery by LS brine injection should increase with increasing $\alpha$ (i.e.\ for networks with larger fractions of OW pores). For networks with large $\alpha$ values, the imbibition phase is shorter, and more wettability modification should therefore take place during the drainage phase. This should enhance the prospect of improved microscopic sweep efficiency in the OW pores and could potentially lead to an overall increase in oil recovery. Networks with small $\alpha$ values, on the other hand, would have less potential for improved microscopic sweep efficiency because a larger fraction of OW pores would be likely to undergo wettability modification before the drainage phase is reached. Note that, for these theoretical considerations, we are again assuming that the LS-induced contact angle reduction does not cause the OW pores to become WW. The results presented in Section~\ref{Results} were all generated using networks with identical OW and WW pore fractions. Therefore, to test the theory outlined above, we have performed additional simulations (not shown) for networks with different $\alpha$ values (note that these networks still have spanning OW and WW clusters, and that all other parameters remain at their Section~\ref{SecLSW} base case values). For $\alpha = 0.465$, we find that LS brine injection displaces 2.5\% more of the resident oil than HS brine injection, while, for $\alpha = 0.535$, we find that LS brine injection displaces 9.1\% more of the resident oil than HS brine injection. Recall that, in the base case simulation ($\alpha = 0.5$), LS brine injection displaced 4.9\% more of the resident oil than HS brine injection. These results show a clear trend of increasing incremental oil production by LS brine injection with increasing $\alpha$, which is entirely consistent with our theoretical underpinnings.

\begin{figure}
\subfloat[\label{SwitchFiga}]{{\includegraphics[width=5.2cm]{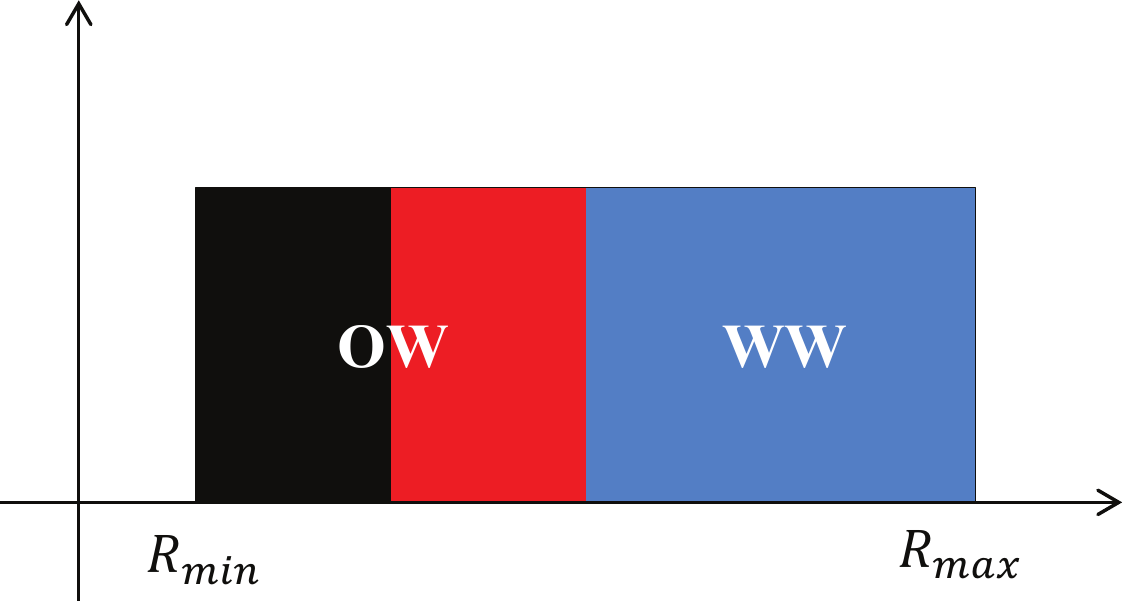}}}
\quad
\subfloat[\label{SwitchFigb}]{{\includegraphics[width=5.2cm]{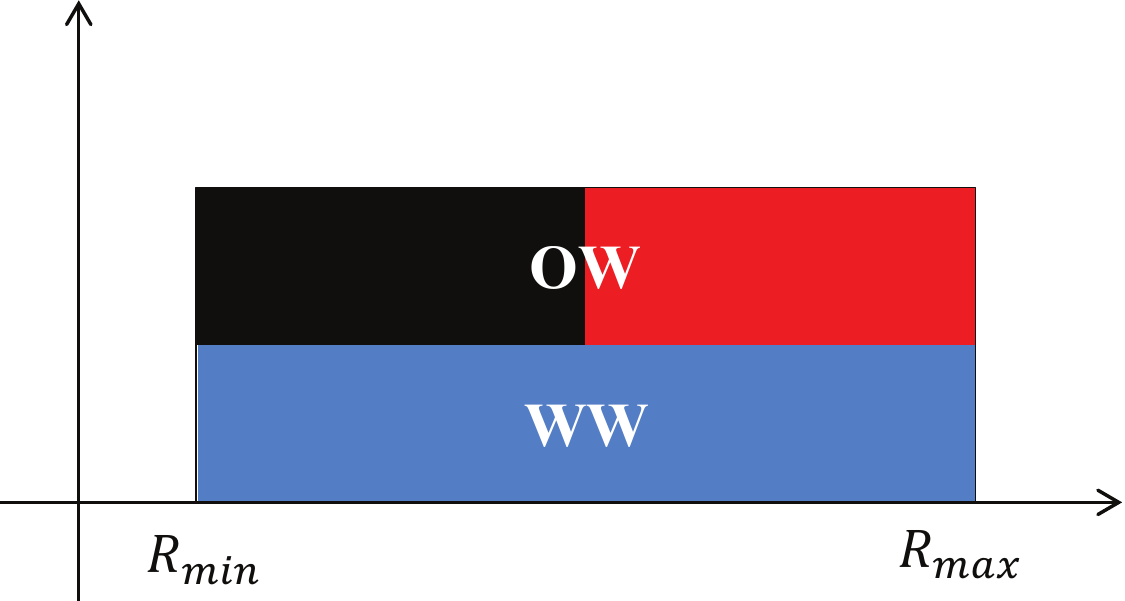}}}
\quad
\subfloat[\label{SwitchFigc}]{{\includegraphics[width=5.2cm]{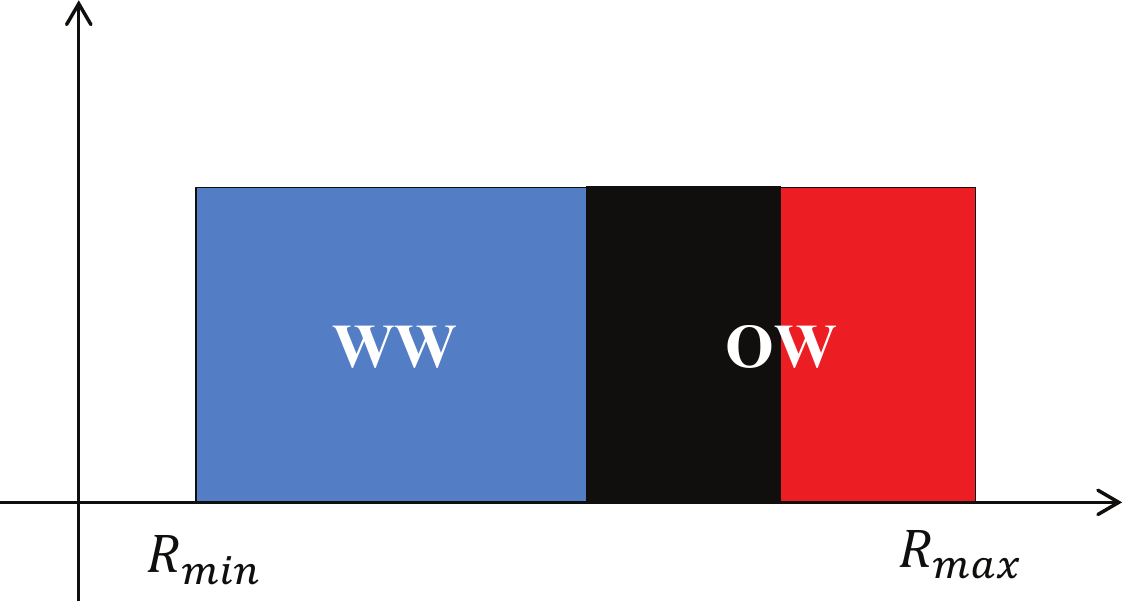}}}
\\
\subfloat[\label{SwitchFigd}]{{\includegraphics[width=5.2cm]{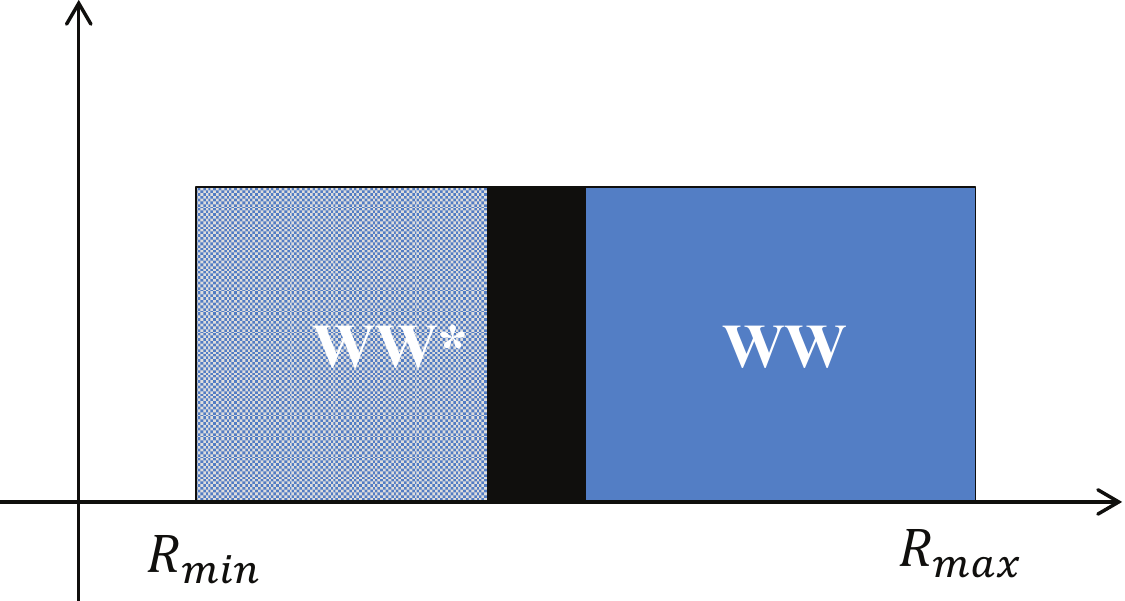}}}
\quad
\subfloat[\label{SwitchFige}]{{\includegraphics[width=5.2cm]{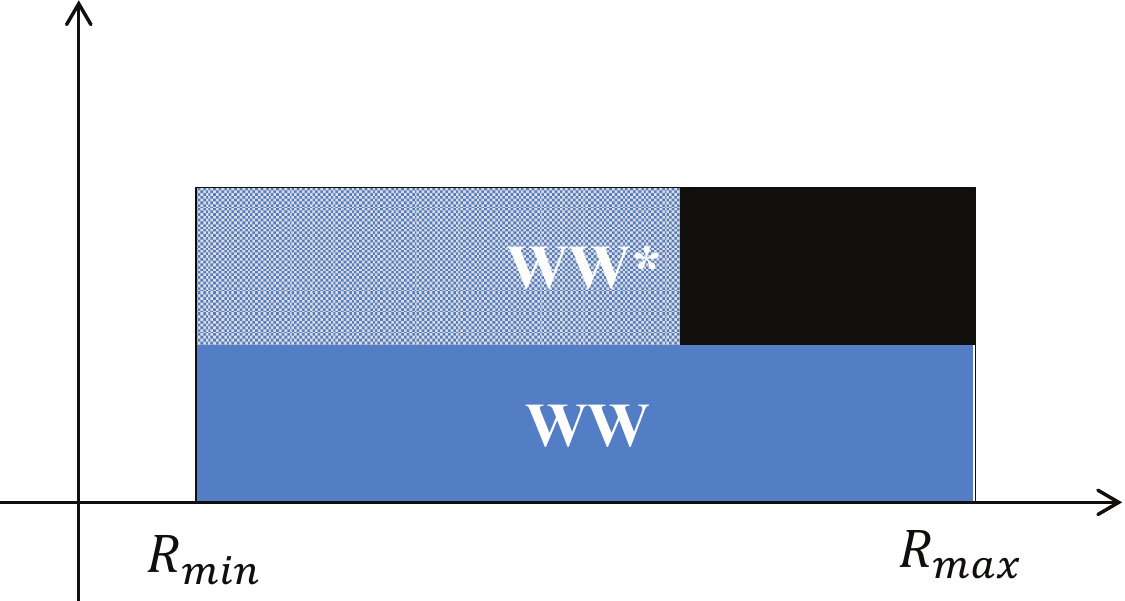}}}
\quad
\subfloat[\label{SwitchFigf}]{{\includegraphics[width=5.2cm]{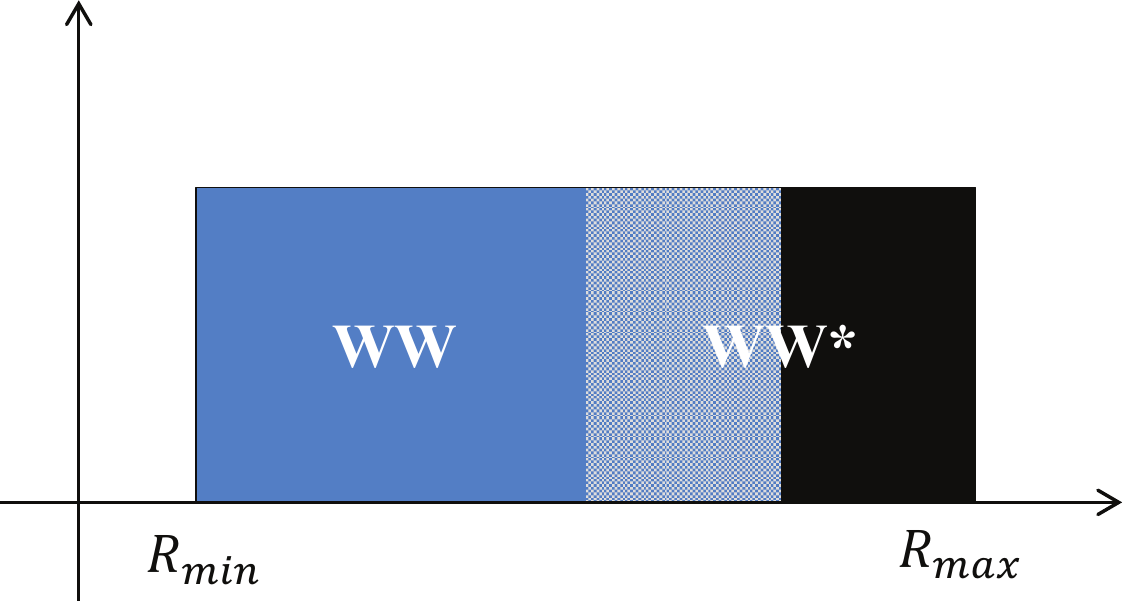}}}
\\
\subfloat[\label{SwitchFigg}]{{\includegraphics[width=5.2cm]{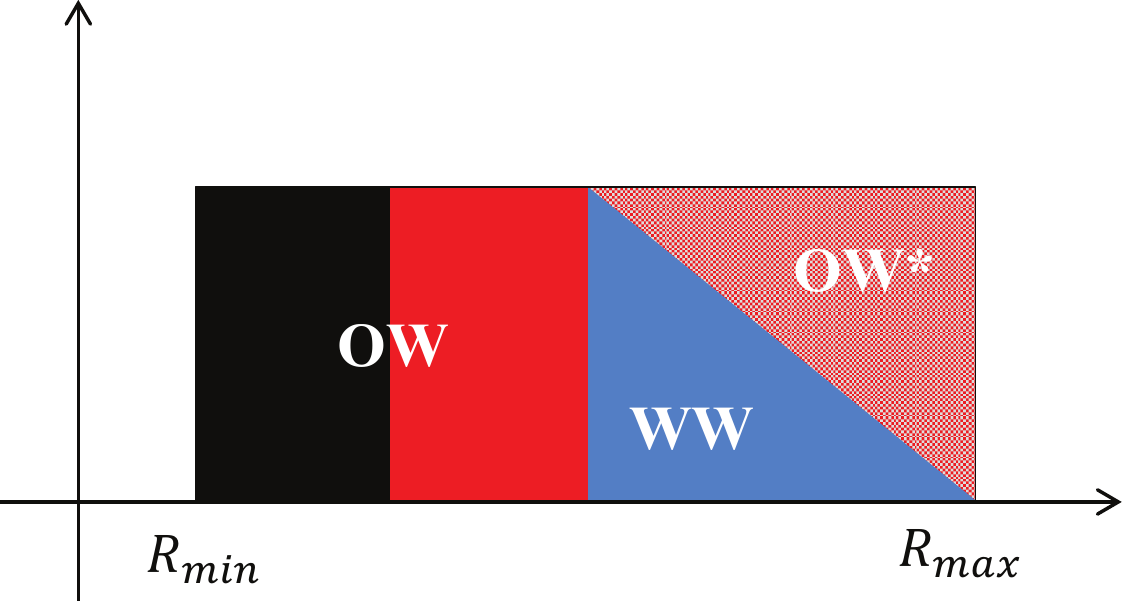}}}
\quad
\subfloat[\label{SwitchFigh}]{{\includegraphics[width=5.2cm]{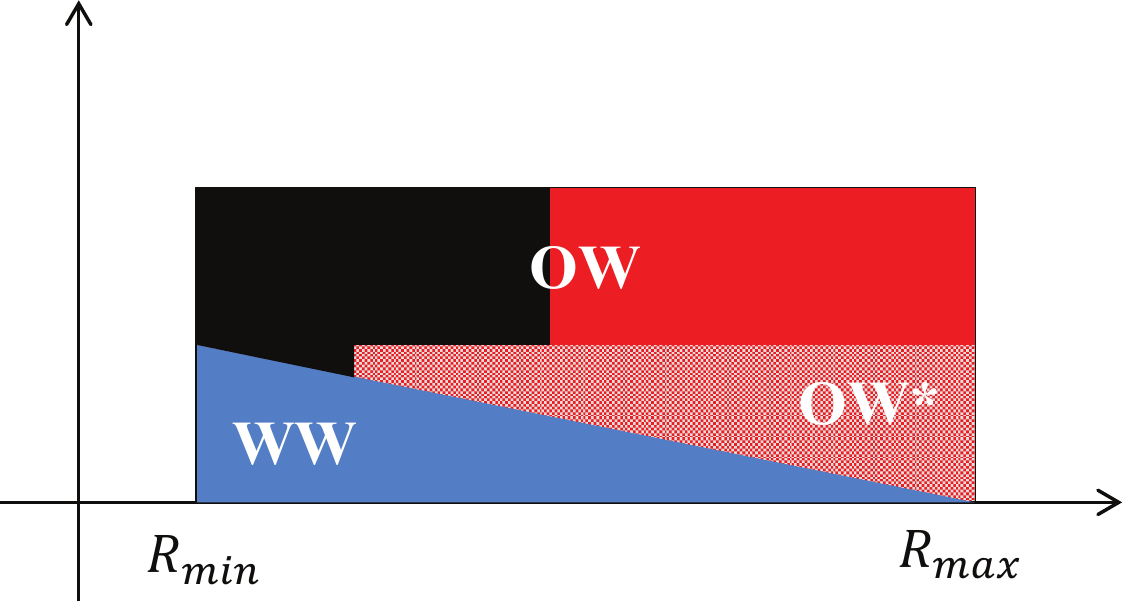}}}
\quad
\subfloat[\label{SwitchFigi}]{{\includegraphics[width=5.2cm]{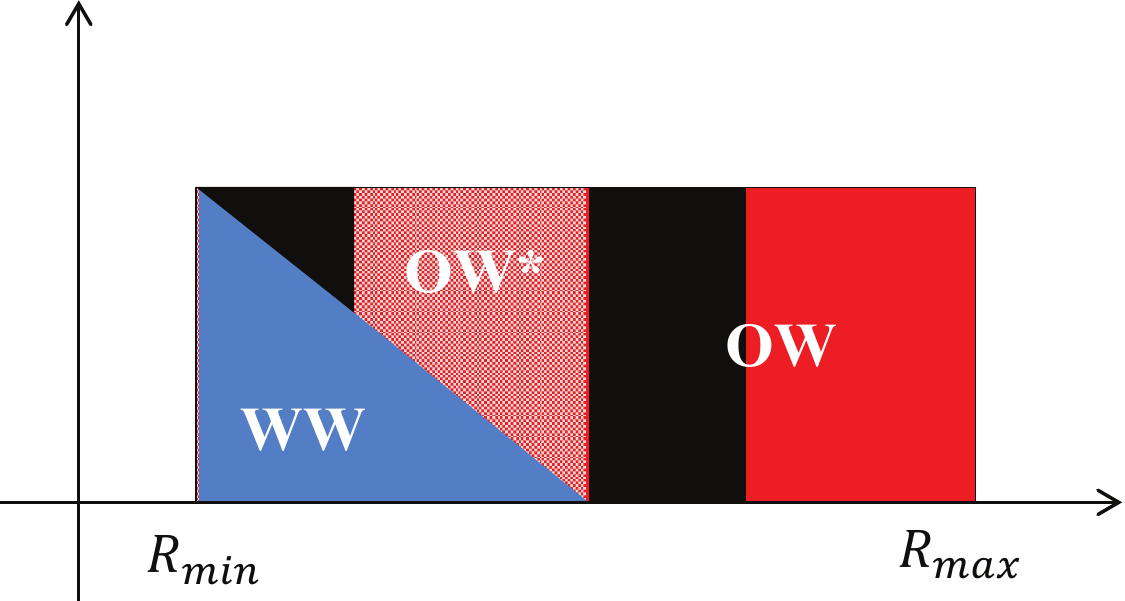}}}
\caption{Schematic plots showing idealised depictions of potential outcomes from representative simulations of (a-c) HS brine injection, (d-f) LS brine injection where OW pores can become weakly WW (here denoted WW$^*$) and (g-i) LS brine injection where WW pores can become weakly OW (here denoted OW$^*$). Plots are based on \textit{in silico} observations of pore-filling sequences for (a, d, g) MWS, (b, e, h) FW and (c, f, i) MWL networks initialised with 50\% WW pores and 50\% OW pores. The blacked-out area on each plot denotes the subset of oil-filled pores that could be expected to be trapped in each case. See the main text for a detailed explanation of the features of the various plots. \label{SwitchFig}}
\end{figure}

We shall now briefly consider cases where LS brine injection causes the contact angles of the WW or OW pores in a non-uniformly wetted network to cross 90{\degree} (such that the network becomes uniformly wetted over the course of the displacement). We emphasise from the outset of this section that, while we do not report any explicit numerical results, the theoretical arguments that we present are all based on observations from an extensive model sensitivity analysis. Model simulations indicate that, when pores can dynamically change their wetting \textit{class}, the sequence of pore filling (and extent of any LSE) in a given simulation is strongly dependent upon the particular distribution of initial and modified pore contact angles (amongst other factors). Therefore, to present results from just one or two simulations would be misleading about the wider qualitative trends that we observe, while to present results from a full suite of simulations would be disproportionate. For the scenarios that we have discussed to this point, the wettability class of the network (MWS, FW or MWL) has had no real consequence for the qualitative outcome of LS brine injection. The situation is rather different, however, when LS brine causes a subset of pores to change their wetting preference. To fully appreciate the mechanisms by which LS brine injection may or may not improve oil recovery under this assumption, it is instructive to first consider the qualitative outcome of standard HS brine injection in networks of each wettability class. Assuming, for illustrative purposes, that $\alpha = 0.5$, Figures~\ref{SwitchFiga} to \ref{SwitchFigc} show schematic diagrams of the groups of pores that we would expect to be displaced by HS brine injection for MWS, FW and MWL networks, respectively. The oil-filled pores that would be trapped during the flood are shown in black and, in each case, represent the \textit{smallest} OW pores in the network (i.e.\ those with the most negative capillary entry pressures).

Consider first the scenario where LS brine injection dynamically reduces pore contact angles and causes the initially OW pores in the network to become WW (as has been directly observed in the experimental study of \citet{Khis17}). For these initially OW pores, the largest-to-smallest drainage filling sequence would no longer occur, and the pores would instead be \textit{imbibed} in a smallest-to-largest filling sequence. Consequently, the \textit{largest} of these oil-filled pores (rather than the smallest) may ultimately become trapped. Hence, it appears that there would be no explicit \textit{volumetric} benefit to this process and any improvement in oil recovery by LS brine injection would likely require an overall increase in microscopic sweep efficiency. High microscopic sweep efficiency could be guaranteed in this case by delaying the injection of LS brine until the drainage phase has begun (i.e.\ once $P_c < 0$). All newly WW pores would then immediately satisfy their (positive) entry pressures, and much of the remaining oil would likely be recovered regardless of the precise network properties. However, if LS brine was introduced into the network at some positive $P_c$ (and the $P_c$ remained positive thereafter due to subsequent contact angle modification), the chances of an improvement in microscopic sweep efficiency would be strongly dependent upon the features of the pore space.

To improve the microscopic sweep efficiency by LS brine injection in this case, it is essential that a subset of the \textit{newly} WW pores can realise capillary entry pressures that are larger than the prevailing $P_c$ and thereby undergo immediate imbibition. It is complex to predict the frequency with which this may occur for a given \textit{in silico} network (depending as it does upon the timing of the changes in wetting preference, the PSD of the network and the instantaneous distributions of the initially WW and newly WW pore contact angles), but it is possible to broadly categorise the different wettability classes in terms of their likely potential for improved microscopic sweep efficiency by LS brine injection. In a MWL network, for example, any newly WW pore would be \textit{larger} than all initially WW pores. Therefore, to undergo immediate imbibition, a newly WW pore would have to become \textit{more WW} than at least some of the initially WW pores. In contrast, for a MWS network, any newly WW pore would be \textit{smaller} than all initially WW pores. Therefore, the immediate imbibition of a newly WW pore in a MWS network could potentially occur even if it was \textit{less WW} than some of the initially WW pores. This condition seems far less stringent than the corresponding condition for MWL networks, and we therefore envisage that MWS networks would have greater potential for improved microscopic sweep efficiency than MWL networks in this case. Since FW networks initially contain WW pores and OW pores of \textit{all} sizes, we also envisage that the potential for improved microscopic sweep efficiency by LS brine injection in FW networks would lie somewhere between that for MWS and MWL networks in this case. Based on these theoretical considerations, Figures~\ref{SwitchFigd} to \ref{SwitchFigf} show schematic diagrams of hypothetical outcomes from LS brine injection into idealised MWS, FW and MWL networks (note that WW$^*$ denotes \textit{newly} WW pores). The MWL diagram assumes no increase in microscopic sweep efficiency, the FW diagram assumes a small increase, and the MWS case assumes a slightly larger increase. Note that the trapped oil pores all have larger radii than those in the equivalent HS cases (Figures~\ref{SwitchFiga} to \ref{SwitchFigc}). Hence, in this scenario (where some OW pores can become WW following exposure to LS brine), even an increase in microscopic sweep efficiency would provide no guarantee of an overall improvement in oil production.

Finally, we consider the scenario where the contact angles in a non-uniformly wetted network are dynamically increased by invading LS brine and the initially WW pores change their wetting preference to become OW. Note that LS brine injection cannot improve microscopic sweep efficiency in this case (increased oil trapping is, in fact, more likely) and any improvement in oil recovery requires a pore filling sequence that is more volumetrically favourable than the sequence for a standard waterflood (c.f.\ the \textit{``pore sequence effect''}). An immediate consequence of this condition is that MWS networks --- the most favourable wettability class in the above scenario --- have \textit{no} potential for improved oil recovery when injected LS brine causes WW pores to become OW. As shown by Figure~\ref{SwitchFiga}, a standard waterflood in a MWS network will tend to trap only the very smallest oil-filled pores and this outcome cannot therefore be bettered by simply altering the pore filling sequence. In general, for the case that LS brine injection causes WW pores to switch their wetting preference, production of incremental oil requires some of the larger initially OW pores (which would otherwise be trapped) to be recovered at the expense of some of the smaller newly OW pores (which would otherwise be displaced by imbibition). The greater the number of pores for which this phenomenon occurs, the greater the potential volumetric benefit to overall oil production. Hence, the optimum approach in this scenario would be to introduce LS brine into the network at the earliest possible stage to maximise the number of small WW pores that become OW.

As for the above scenario of OW pores becoming WW, the efficacy of an approach where LS brine injection causes WW pores to become OW is difficult to predict \textit{a priori}. It is possible to assert, however, that the potential for the displacement of large, initially OW pores at the expense of small, newly OW pores should be greater in MWL networks than in similar FW networks. Unlike for a FW network, in a MWL network the initially OW pores will \textit{all} be larger than any newly OW pore. Given that the OW pores in a network will tend to fill in a largest to smallest sequence, the MWL wettability class therefore appears to maximise the chances for oil trapping in small, newly OW pores. Figures~\ref{SwitchFigg} to \ref{SwitchFigi} show schematic diagrams of hypothetical outcomes from LS brine injection in networks of each wettability class, where the newly OW pores are denoted OW$^*$ (for illustrative purposes, we assume that 50\% of the initial WW pores become OW during the imbibition phase). Based on the above theoretical considerations, the MWS case exhibits no overall change to the filling sequence (Figure~\ref{SwitchFigg}), the FW case exhibits a small change (Figure~\ref{SwitchFigh}) and the MWL case demonstrates a larger change (Figure~\ref{SwitchFigi}). Figures~\ref{SwitchFigh} and \ref{SwitchFigi} both suggest that the oil recoveries for LS brine injection would exceed those obtained from equivalent HS brine displacements (Figures~\ref{SwitchFigb} and \ref{SwitchFigc}, respectively). Note, however, that these plots do not account for the increase in topological oil trapping that is typically observed in simulations where injected LS brine dynamically increases contact angles. Hence, the likely outcome for LS brine injection in a given network is not clear-cut, and both increases and decreases in overall oil recovery could again be possible in this scenario.

\section{Conclusions \label{Conclu}}
It is important to emphasise that we have taken a relatively modest approach to our investigation of the LSE in this study. We have utilised a steady-state fluid displacement model in which corner flow is neglected, and where the evolution of brine salinity in the pore space is estimated using a dynamic tracer injection algorithm. The model also neglects any explicit pore-scale chemistry and assumes that LS brine injection leads to a direct alteration of local wettability. However, this pragmatic approach, combined with a focus on idealised \textit{in silico} pore structures, has made it feasible to identify several factors that may be critical to the LSE in non-uniformly wetted networks. In line with experimental observations, our simulations and analysis demonstrate that LS brine injection can have a positive, negative or neutral effect on overall oil recovery.

For non-uniformly wetted networks in which LS brine injection cannot alter the actual wetting \textit{class} of individual pores (only the strength of the wetting preference), our results have indicated that a dynamic, LS-induced reduction in contact angles can significantly increase the production of oil. We have shown that the OW pores provide the only viable source of incremental oil in non-uniformly wetted networks and, consequently, that oil recovery by LS brine injection is optimised when wettability modification occurs primarily during the drainage cycle. Simulations demonstrate that the fraction of OW pores in the network $\alpha$, the average network connectivity $\bar{Z}$, and the initial HS water saturation $S_{wi}$ can all strongly influence the degree to which LS-induced wettability modification takes place during drainage. Hence, we conclude that all three of these parameters can play a critical role in determining the extent of any positive LSE in non-uniformly wetted porous media.

The results reported in this study demonstrate that the factors that promote incremental oil production by LS brine injection can be distinctly different for uniformly wetted and non-uniformly wetted pore networks. These findings, which can help to clarify the prevailing uncertainty in the experimental literature, further underline the importance of identifying the initial and final wettability states when assessing coreflood performance in LS waterflooding studies.

\appendix*
\section{Dynamic salinity updates in flowing water-filled pores}
For each post-breakthrough $P_c$ reduction that permits the ingress of LS brine, a period of convective tracer transport is simulated to determine the new salinity distribution in the flowing water. Tracer of concentration one (i.e.\ salinity = 0) is introduced at the network inlet and, at each timestep $\Delta t$, tracer concentrations $C_{old}$ of flowing water-filled pores are updated according to their elemental flow rates $q$ and volumes $V$ as follows:
\begin{enumerate}[label=(\roman*)]
\item calculate the mass of tracer that leaves each pore, $M_{out} = q \cdot \Delta t \cdot C_{old}$;
\item calculate the total flow $q_{node}$ and total mass of tracer $M_{node}$ that enters each node by summing the $q$ and $M_{out}$ values from immediate upstream pores;
\item assume perfect mixing in each node and calculate the mass of tracer that enters each immediate downstream pore, $M_{in} = \left(q / q_{node}\right) \cdot M_{node}$;
\item calculate the new tracer concentration in each pore, $C_{new} = C_{old} + (M_{in} - M_{out}) / V$.
\end{enumerate}
At each timestep of the above procedure, the mass of tracer that leaves a pore cannot exceed the initial mass of tracer in the pore (i.e.\ $M_{out} \leqslant C_{old} \cdot V$). The timestep length is therefore given by $\Delta t = \min\left(V/q\right)$, where the minimum is taken over all flowing water-filled pores.

\begin{acknowledgments}
The authors would like to thank TOTAL S.A.\ for their invaluable technical support and financial assistance.
\end{acknowledgments}

\bibliography{Manuscript_Watson_revised}

\end{document}